\documentclass[rmp,aps,nofootinbib,twocolumn,floatfix,showpacs]{revtex4}

\usepackage{graphics}
\usepackage{epsfig}
\usepackage{bm}

\def\he4{$^4$He}
\def\hee3{$^3$He}
\def\Am3{\AA$^{-3}$}
\def\beq{\begin{equation}}
\def\eeq{\end{equation}}

\begin{document}

\title{Colloquium: Supersolids: What and Where are they ?}

\author{Massimo Boninsegni}
\email{m.boninsegni@ualberta.ca}
\affiliation{Department of Physics, University of Alberta. Edmonton, Alberta T6G 2E1, Canada}

\author{Nikolay V. Prokof'ev}
\email{prokofev@physics.umass.edu}
\affiliation{
Department of Physics, University of Massachusetts, Amherst, MA 01003, USA
}

\include{revmacro}
\date{\today}

\begin{abstract}
The ongoing experimental and theoretical effort aimed at understanding non-classical rotational inertia in solid helium, has sparked renewed interest in the supersolid phase of matter, its microscopic origin and character, and its experimental detection. The purpose of this colloquium is {\it a}) to provide a general theoretical framework for the phenomenon of supersolidity and {\it b}) to review some of the experimental evidence for solid $^4$He, and discuss  its possible interpretation in terms of physical effects underlain by extended defects (such as dislocations). We provide quantitative support to our theoretical scenarios by means of first principle numerical simulations. We also discuss alternate avenues for the observation of the supersolid phase, not involving helium but rather assemblies of ultracold atoms.

\end{abstract}

\maketitle
\tableofcontents

\section{Introduction}
\label{sec:intro}

The search for exotic phases of matter drives much of the fundamental research in condensed matter physics. Of particular interest are phases displaying simultaneously different types of order. The archetypal example is the so-called {\it supersolid}, an intriguing,  conjectured macroscopic manifestation of quantum mechanics, which has until now eluded direct, unambiguous experimental observation. In recent times, the search for this fascinating phase of matter has focused on solid helium, but alternate avenues to its possible observation are emerging, notably in the area of cold atom physics.
\\ \indent
At the time of this writing,  consensus is still lacking, as to whether the wealth of experimental evidence accumulated  for solid helium (especially over the past eight years), indisputably  points to the observation of the supersolid phase of matter. Among the various proposed theoretical scenarios, none appears capable of accounting for all the puzzling, often conflicting experimental findings \cite{Balibarrec}.
\\ \indent
In this Colloquium, we  make no attempt to offer the final word on this subject. Rather, we  modestly limit ourselves to illustrating {\it one} theoretical interpretation of {\it some} of the existing helium phenomenology. Specifically, we shall describe a theoretical scenario in which extended defects, such as dislocations and grain boundaries, play a crucial role. Such a scenario is born out of, and supported by, first principle microscopic calculations, based on Quantum Monte Carlo techniques.
\\ \indent
Motivated by the impetuous progress made in the  context of ultracold atom physics,  in light of the clear promise that such a field holds toward the realization of artificial quantum many-body systems, we discuss the possible observation of the supersolid phase in ultracold atom assemblies. In this case too, our theoretical proposals are largely based on numerical evidence from computer simulations.

\subsection{Basic definitions}
What is a ``supersolid" ? As it turns out, this concept defies in part one's intuition. Although a rigorous theoretical definition can be provided, it can be tricky  at times to establish unambiguously whether a given physical system meets all the requirements that warrant such a denomination. For this reason, we shall first proceed to state those conditions as clearly as we can, in order to set up the framework for the rest of our discussion.
\\ \indent
Taken literally, of course, the word {\it supersolid} means ``superfluid solid". This implies that a supersolid substance ought display the same basic qualities of a conventional solid, {\it as well as} those of a superfluid. This poses an intellectual challenge right away. For, one is hard put forming a mental picture of how the very different, and seemingly antithetical solid and superfluid physical behaviors, could be merged into a single homogeneous phase of matter, at least  based on familiar {\it macroscopic} notions. Should one imagine a supersolid to be a peculiar solid, capable of flowing just like a liquid (and in addition, without dissipation)~? Or, conversely, something more like a ``stiff liquid", resisting  shear like a solid~? While there are some elements of truth in these tentative descriptions, a rigorous, consistent definition of the supersolid phase is best achieved by taking on a microscopic approach, supplemented by the theoretical notion of {\it long-range order}.

In order to keep the discussion as simple and transparent as possible, we shall for the moment restrict it to
three-dimensional (3D) systems. In this case, both solid and superfluid phases have well-defined order parameters that characterize them. In lower
dimensional systems, solid and superfluid can be
defined {\it without} reference to order parameters (any such quantity vanishes identically at any finite temperature). Instead, one must speak of long-range (non-integrable)
power-law correlations in the order parameter {\it field} \cite{KT}. This
difference, however, is not of fundamental importance to
our presentation, since modifications of theory required to deal with  reduced dimensionality are relatively simple, and well understood.

\subsubsection{Order in a solid}
In a {\it crystalline} solid, elementary constituents (i.e., atoms or molecules) are spatially arranged in an orderly fashion, each occupying one of a discrete set of well-defined sites of a three-dimensional periodic lattice\footnote{We assume for simplicity a {\it Bravais} lattice.}. In order to make a more formal statement, let us introduce the (time-averaged\footnote{In any real crystal, thermal and quantum fluctuations cause the local density to deviate instantaneously from its time-averaged value.}) local  density $\rho({\bf r})$ of particles forming the crystal. The average value of $\rho({\bf r})$ in a macroscopic sample of crystal of volume $\Omega$ is given by
\begin{equation}
\bar\rho \equiv\frac{1}{\Omega}\ \int d^3r\ \rho({\bf r}).
\end{equation}
Let $\delta\rho({\bf r})\equiv\rho({\bf r})-\bar\rho$ be the local deviation of the density from the sample-averaged value $\bar\rho$.  In a system that does not break translation invariance, such as a gas or a liquid, $\rho({\bf r})=\bar\rho$ and  $\delta\rho({\bf r}) = 0$. In a crystal, on the other hand, translational invariance is broken, i.e., $\delta\rho({\bf r})$ does not vanish identically, and  ordering is expressed through the following condition:
\beq\label{period}
\delta\rho({\bf r}) = \delta\rho({\bf r}+{\bf T}),
\eeq
holding for {\it any}  vector {\bf T} belonging to the discrete set of lattice vectors (i.e., vectors connecting any two lattice sites) defining the crystal. Consider the three-dimensional Fourier transform of $\delta\rho({\bf r})$
\begin{equation}
\tilde \rho({\bf k}) = \frac {1}{\Omega}\ \int d^3r\ \delta\rho({\bf r})\ e^{-i{\bf k}\cdot{\bf r}}
\end{equation}
Condition (\ref{period}) implies that $\tilde \rho({\bf k})$,
and its squared magnitude $S({\bf k})$, known as the {\it Static Structure Factor}, display peaks in correspondence of
wave vectors {\bf G} in the so-called {\it reciprocal lattice}, i.e., fulfilling the condition
\beq
{\bf G}\cdot{\bf T} = 2\pi n
\eeq
for any vector {\bf T} defined as above, with $n$ being an integer. These peaks in $S({\bf k})$ manifest themselves experimentally as maxima in the intensity of light scattered off a crystalline sample, at specific angles  \cite{chaikin}. Because condition (\ref{period}) is assumed to hold for a macroscopic sample, one speaks of {\it density long-range order} (LRO).
\\ \indent
Broken translation invariance has the immediate {\it macroscopic} consequence of imparting to the system the well-known property of resistance to shear, which a fluid does not possess. It is important to note that, while LRO implies broken translation invariance, the converse is not true.
Glasses, for example, are substances which share many of the macroscopic properties of crystalline solids, in which  translation invariance is broken, but $\delta\rho({\bf r})$ fluctuates disorderly and no LRO exists. As we shall discuss later in this chapter, there is in principle nothing preventing some glassy systems from turning superfluid; however, the denomination {\it supersolid} does {\it not} apply here, due to the lack of density LRO. In these cases, one ought more properly speak of {\it superglass}.
\\ \indent
Another point that need be made at the outset, is that condition (\ref{period}) occurs in a solid {\it spontaneously}, exclusively as a result of interactions among elementary constituents, at specific thermodynamic conditions (i.e., temperature and pressure). Such a spontaneous breaking of translation symmetry is an integral part of the definition of supersolid. One could certainly imagine externally imposing a density modulation to a superfluid liquid. For instance, this could be accomplished by adsorbing few layers of $^4$He on a strongly attractive substrate, such as graphite \cite{crowell93}. The local density of successively adsorbed $^4$He layers inevitably reflects the corrugation (i.e., LRO) of the underlying graphite substrate, to lesser a degree for layers increasingly removed from it. Referring to any  superfluid adsorbed $^4$He layer as ``supersolid", on account of the density modulation imposed by a corrugated substrate, is fundamentally incorrect and
constitutes a misnomer.

\subsubsection{Order in a superfluid}
{\it Superfluidity} is the property of a substance sustaining persistent, dissipation-less flow. It was first observed in the liquid phase of the most abundant isotope of helium ($^4$He), independently by Kapitsa \cite{kapitza} and Allen and Meisener \cite{allen} in 1938. Its fundamental relevance extends  considerably beyond the physics of condensed helium. For example, it is now widely  accepted that {\it superconductors} are essentially charged superfluids \cite{tilley}.
\\ \indent
A useful phenomenological model introduced by Tisza shortly after the discovery of superfluidity in liquid \he4 \cite{tisza},  formally expresses the local density $\rho({\bf r})$ of a superfluid system, as a sum of two contributions:
\beq
\rho({\bf r}) = \rho_S({\bf r}) + \rho_N({\bf r})
\eeq
where the subscripts $S$ and $N$ stand for ``superfluid" and ``normal" component. The superfluid component carries no entropy and can sustain flow indefinitely, whereas the normal component is subjected to dissipation, as any regular fluid. Henceforth, we shall use the notation $\rho_S$ and $\rho_N$ to refer to values averaged over the whole system.  In a translation invariant system, it is $\rho_N({\bf r}) = \rho_N$ and $\rho_S({\bf r})=\rho_S$. Superfluid behavior
typically sets in through  a continuous phase transition at a temperature $T_c$, below which $\rho_S$ becomes nonzero, and increases monotonically as the temperature $T \to 0$. In a translation invariant system, $\rho_S(T=0) = \bar \rho$, i.e., the average density of the system. Phrased differently, the {\it superfluid fraction} $\rho_S/\bar\rho$ approaches unity as $T\to 0$. On the other hand, Leggett showed \cite{Leggett70} that, for any superfluid system breaking translation invariance, it is $\rho_S(T=0) < \bar\rho$.
\\ \indent
A wealth of theoretical and experimental work, spanning now over seven decades, has afforded satisfactory theoretical understanding of the microscopic origin of superfluidity, in $^4$He and other systems.
In particular, it is now understood that superfluidity is a macroscopic manifestation of quantum particles behaving collectively
as classical complex fields \cite{KT,Boris}, and that it
is crucially underlain by a type of quantum statistics (Bose)
since description in terms of classical fields
emerges in the limit of large occupation numbers \cite{Glauber,Langer,Langer2}.

In three dimensional systems,
superfluidity is accompanied by {\it Bose-Einstein condensation} (BEC) \cite{leggett}, a collective phenomenon that takes place at low temperature.
Indeed, there is virtually unanimous agreement that, if superfluidity is to occur in a system of identical particles obeying Fermi statistics,
some physical mechanism must be present, allowing for the formation of {\it pairs}  (e.g., electronic Cooper pairs in superconductors, or atomic pairs in $^3$He \cite{tilley} and cold atomic vapors).
These composite objects act in some sense as bosons, and as such can
undergo BEC\footnote{The relationship between superfluidity and BEC is
straightforward only in 3D; formally BEC is not required for superfluidity,
as evidenced by finite temperature superfluidity in two-dimensional (2D) systems, where BEC
is replaced by topological order and power-law correlations.}.
\\ \indent BEC consists of the occupation of just one quantum-mechanical single-particle state by a {\it finite fraction} of all $N$ particles  in the system\footnote{For simplicity, but with no loss of generality, we restrict our discussion to systems of bosons of spin zero.}. In order to be regarded as such, BEC must take place in a {\it macroscopic} sample of matter; mathematically, this is expressed by taking the {\it thermodynamic} limit, i.e.,  $N, \Omega\to\infty$, but the density $\bar \rho = N/\Omega$ remains finite.
\\ \indent
In a many-particle system that enjoys translation invariance, like a liquid, the single-particle quantum-mechanical state into which particles can condense, is that of a free particle of momentum $\hbar{\bf k}=0$. In order to render the above  statement more quantitative, we introduce the {\it momentum distribution}
\beq\label{nofkdef}
\tilde n({\bf k}) \equiv \langle \hat \psi^\dagger ({\bf k})\ \hat \psi ({\bf k}) \rangle/N
\eeq
where $\langle ...\rangle$ stands for thermal expectation value and $\hat \psi ({\bf k})$, $\hat \psi^\dagger ({\bf k})$ are Bose annihilation and creation operators of a particle of momentum $\hbar{\bf k}$. The operator $\hat \psi^\dagger ({\bf k})\hat \psi ({\bf k})$ is the number operator for particles of said momentum.
\\ \indent
In a Bose condensed system, $\tilde n({\bf k})$ will take the form
\beq\label{nofk}
\tilde n({\bf k}) = n_\circ \delta({\bf k}) + \tilde n_{NC}({\bf k})
\eeq
The first term refers to the condensate, and the quantity $n_\circ$ is referred to as {\it condensate fraction}. It should {\it not} be confused with the {\it superfluid fraction}, namely the fraction of the system that can flow without dissipation. These are two conceptually distinct quantities. In particular, $n_\circ$ is the square of the amplitude of the complex  order parameter for the superfluid. The second term of (\ref{nofk}) represents the contribution to the momentum distribution coming from states of nonzero momentum, none of which is occupied {\it macroscopically} (i.e., by a finite fraction of all the particles in the system).
The condensate fraction approaches a value close to 100\% at low temperature, in a weakly interacting Bose gas. Interactions have the effect of depleting the condensate; for example, the most current experimental estimate of $n_\circ$ in liquid $^4$He at zero temperature and at saturated vapor pressure, is about 7.5\% \cite{glyde00}.
\\ \indent
In order to elucidate how a finite condensate fraction corresponds to a kind of order, consider the Fourier transform of the momentum distribution, known as {\it one-particle density matrix}
\beq
n({\bf r},{\bf r^\prime}) =\langle \hat\psi^\dagger({\bf r})\hat\psi({\bf r^\prime})\rangle
\eeq
Here, the field operators $\hat\psi({\bf r})$ and $\hat\psi^\dagger({\bf r})$ annihilate and create a particle at position {\bf r}. They are
Fourier transforms of $\hat \psi ({\bf k})$ and $\hat \psi^\dagger ({\bf k})$.

In a system enjoying translation invariance, such as a liquid or a gas, it is 
$n({\bf r},{\bf r^\prime}) = n({\bf r}-{\bf r^\prime})$. 
If translation invariance is broken, as in a crystal, it is customary to consider, for simplicity, the spatially averaged function
\begin{equation}
n({\bf r}) = \frac{1}{\Omega} \int d^3r'\,
n({\bf r}', {\bf r}'+{\bf r}) \;.
\end{equation}
It can be easily shown that condition (\ref{nofk}) is equivalent to
\begin{equation}
n(r) \to n_{\circ} \; {\rm as} \; r \to \infty \;,
\label{odlro}
\end{equation}
implying that a state of a system comprising a thermodynamic number of particles, in which a particle is removed at a given position {\bf r}, has a finite quantum-mechanical amplitude over a state in which an identical particle is removed at an arbitrarily large distance away from {\bf r}. Condition (\ref{odlro}) expresses the presence of a type of order, referred to as {\it off-diagonal long range order} (ODLRO).
\\ \indent
One way to come to terms with the mind-boggling notion of ODLRO, consists of accepting that, because particles are indistinguishable and can trade place with one another, each particle can be seen as being essentially delocalized throughout the whole system \cite{Leggett70,Kohn}.

\subsubsection{Order in a Supersolid}
Henceforth, we shall adopt the term ``supersolid", to refer to  a {\it homogeneous}\footnote{As opposed to simple {\it coexistence} of two phases, each featuring only one kind of order.} phase of matter in which both LRO and ODLRO, as defined above, exist simultaneously and appear spontaneously for the same species of particles. The definition of ODLRO is the same in a system featuring crystalline order. Conventional momentum is replaced by {\it crystal momentum}, but BEC, if it occurs,  still takes place in the ${\mathbf k} = 0$ state (up to a reciprocal lattice vector), by symmetry.   Moreover, since a crystal breaks translation invariance the superfluid fraction ought to saturate to a value  less
than unity as $T\to 0$.
\\ \indent
At first, one might think that the intrinsic localization of particles characterizing a crystal phase, and the delocalization ensuing from the condition (\ref{odlro}) (as well as superfluidity) ought to be incompatible. Indeed, this was the conclusion reached by Penrose and Onsager, arguably the first to pose the question of the existence of a supersolid in 1956. In their seminal paper, they contended that no ODLRO can exist in a crystalline solid. Their argument
is based on a variational model of a crystal with atoms  ``pinned" at equilibrium lattice positions \cite{Penrose}.  The statement is that particle localization, due to crystallization or other causes (e.g., disorder), acts so as to prevent macroscopic quantum-mechanical exchanges of indistinguishable particles, thereby {\it de facto} removing all effects of quantum statistics (including BEC).
\\ \indent While seemingly plausible, and in fact re-iterated even many years later \cite{nozieres}, the argument of Penrose and Onsager is no formal proof of the non-existence of a supersolid. Indeed, shortly after the publication of the article by Penrose and Onsager,  Gross \cite{Gross,Gross2} showed that a superfluid system described by the non-linear classical field 
equation may feature a density wave modulation. This finding, which qualifies as the first
theory of the supersolid phase, was essentially overlooked, in part due to
incomplete understanding of the conditions leading to the validity of the calculation by Gross.
In 1962, C. N. Yang \cite{Yang} proposed that in a crystal of helium, in which atoms enjoy a high degree of quantum delocalization, exchanges of adjacent atoms might in fact be significant. Successively, Leggett \cite{Leggett70} revisited Yang's argument, suggesting an experiment in which the (presumably small) superfluid response of a $^4$He crystal could be measured.
\\ \indent
A different, intriguing scenario for the occurrence of supersolid order in a highly quantum crystal, came from Andreev and Lifshitz \cite{Andreev69}, and Chester \cite{Chester}, and is based on the predicted high mobility, at low temperature, of point defects like vacancies or interstitials, which are thermally activated in most materials, but may be present even in the ground state of such quantum many-body systems as $^4$He. These delocalized defects are often
referred to as ``zero-point defects", to emphasize the notion  that they are not
produced by deliberately removing (adding)  atoms from (to) the lattice. Instead,
the lattice period and the unit cell volume, $\Omega_0$, are
automatically adjusted to have a non-integer number of particles
per unit cell, $\bar {\rho } \Omega_0$, even in the fully equilibrated state.
Conceivably, a gas of such {\it repulsively} interacting  point defects ought undergo BEC, and turn superfluid, at low temperature.
More recently, it was pointed out that the ``zero-point defects"
supersolid scenario is the only one possible in perfect continuous space
crystals \cite{PS2005} because superfluid states are necessarily gapless with respect to adding and removing particles to/from the system. Consequently,
integer $\bar {\rho } \Omega_0$ in a supersolid may occur only as a zero-measure
coincidence.
\\
A crystal of $^4$He has always been regarded as the most likely candidate to display supersolidity, arising through any of the above scenarios, even though no reliable quantitative predictions have ever been furnished  of the magnitude of the effect, or the transition temperature. This is perhaps the main reason why, despite an intense experimental effort,  no unambiguous experimental observation of supersolid behavior in $^4$He was reported for decades, following the original theoretical predictions. In 1992, a rather sobering assessment of the prospects of
successful observation of a supersolid phase of helium, was offered by Meisel \cite{Meisel}.
\\ \indent
An exciting, unexpected turn of events took place in 2004, with the discovery, by
Kim and Chan, of a downward shift in the period of a torsional oscillator filled with
solid $^4$He, at temperatures below 250 mK \cite{KC,KC2}.
The interpretation of these findings in terms of a superfluid mass decoupling
within the helium solid unquestionably appears legitimate, serious and plausible.
However, as we shall discuss in the next sections, the final word is yet
to be spoken as to whether the effect first reported by Kim and Chan, and since then confirmed by a number of other groups, indeed
signals the onset of a new phase of matter.
\\ \indent
Irrespective of the above, though, there is no question that the discovery
has sparked a renewed interest in the phenomenon of supersolidity,
from which novel, interesting lines of research have spun.
For one thing, achieving a complete, thorough understanding of
the low temperature phase diagram of solid helium has proven a surprisingly complicated,
and surely intriguing problem on its own, both from the theoretical and experimental standpoints. Many
 puzzling, and often apparently conflicting experimental findings, have cast doubts on the
applicability to solid $^4$He of microscopic pictures such as that of
Andreev, Lifshitz and Chester. In particular,
significant attention is being directed to inhomogeneous
scenarios, involving such extended defects as grain boundaries, as well as dislocations.
\\ \indent
Secondly, the search for the supersolid phase of matter has now extended beyond solid helium,
in a new and exciting direction.
The scientific breakthroughs in cold atom physics that have characterized the past
two decades, make it now feasible to investigate
novel phases of matters, not only by providing remarkably clean and controlled experimental
many-particle systems, but also by allowing one to fashion artificial inter-particle potentials,
not arising in any known condensed matter system. The aim here is twofold: on the one hand,
the high degree of control afforded by cold atoms, may render the unambiguous identification of
the sought phase more direct, and easier than in solid helium. More fundamentally, the possibility
of ``tweaking" the inter-particle potential permits one to address basic questions, such
as: which kind of interaction, if any, can underlie a supersolid phase ?

\subsection{Experimental detection}\label{exp}
Because it is a manifestation of quantum mechanics on a macroscopic scale, one would be inclined to assume that
the observation of supersolid  behavior should be relatively straightforward and unequivocal, the only issue being whether the relevant  phenomenology occurs  at  experimental conditions (e.g., temperature) that are presently accessible.
Observation of superfluidity in liquid $^4$He is made relatively simple by the stunning effects that can be produced in laboratory, all of which exploit the unusual flow of the superfluid liquid. The situation is more complicated in the case of a supersolid, which is expected to retain the mechanical properties of a regular solid, notably resistance to shear, preventing it to flow like
a liquid\footnote{Flow arising as a result of {\it plastic} deformation
of a supersolid, is expected to be radically different from that
in a regular solid. The corresponding theory, however, is still
far from being complete (see also below)}. \\ \indent
Strictly speaking, some demonstrations of superfluid behavior, such as persistent flow, or the well-known
{\it fountain effect} \cite{tilley}, should be feasible in a solid as well, with the important caveat that
a supersolid will support dissipation-free flow only
of its own particles; with respect to any other external object, it
will behave as a conventional solid. Clearly, however, the ease and spectacular immediacy of the observation
of such effects,  when they take place in  the liquid superfluid phase, may be difficult to replicate in the solid.
In general,  therefore, the supersolid nature of a condensed matter system can only be typically ascertained by means of relatively indirect observations.

Another potential problem lies in the fact that, as mentioned above, in the low temperature limit neither the superfluid nor the condensate fractions of a supersolid  will approach 100\%, due to interactions among particles and the lack of translation invariance. Rather, they will both saturate to values that could conceivably be very small, rendering the measurement of any ``superfluid signal" problematic.
\subsubsection{Non-classical Rotational Inertia}
\begin{figure}[h]
\includegraphics[width=2.0in]{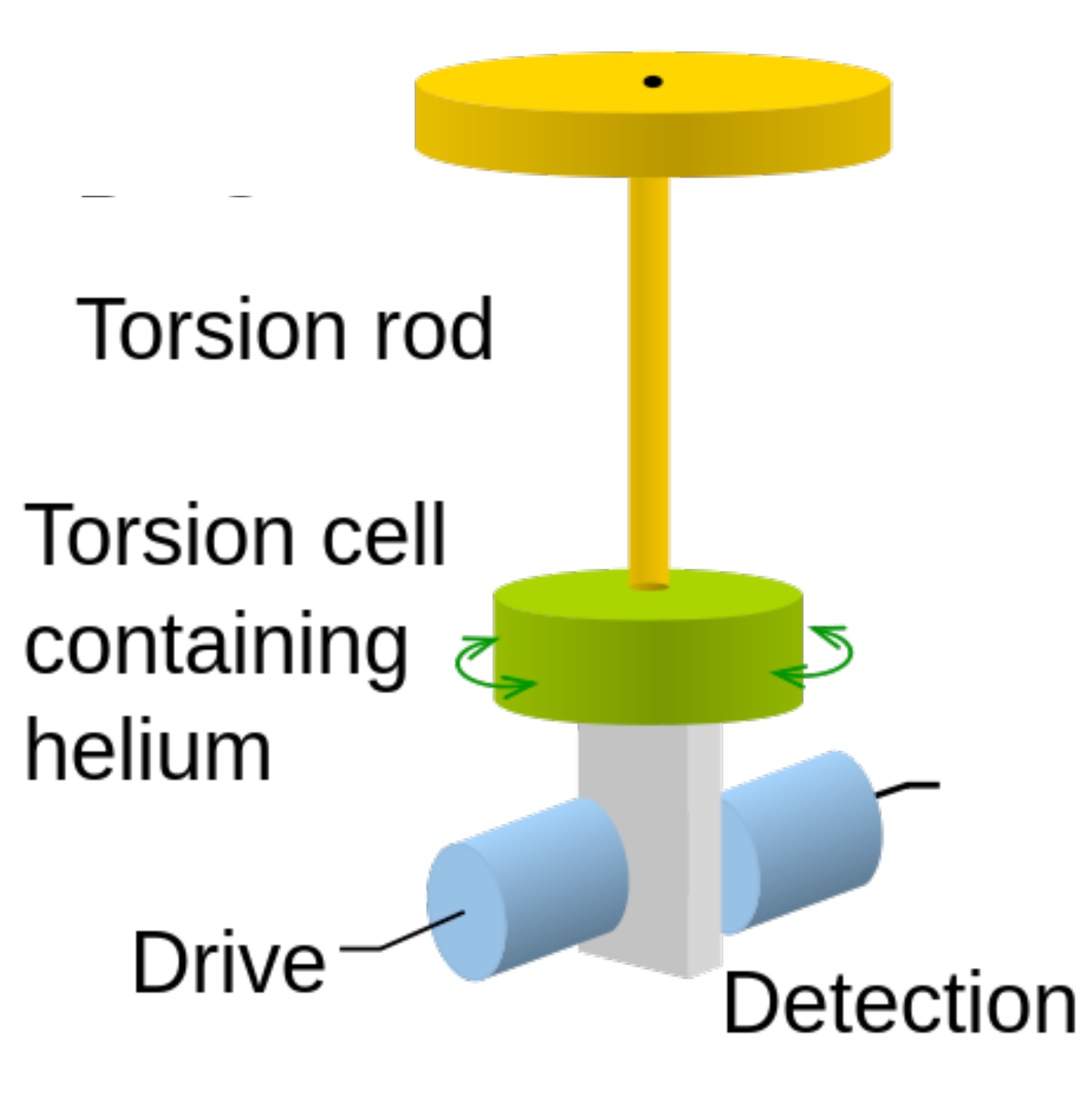}
\caption{Schematic of the experimental Torsional Oscillator set-up aimed at measuring non-classical rotational inertia. The cell is filled with a sample of condensed matter, whose superfluid fraction affects the moment of inertia of the overall system.
}
\label{fI1}
\end{figure}

Perhaps the simplest way of detecting experimentally supersolid behavior,
was suggested by Leggett in 1970 \cite{Leggett70}, and consists of measuring a
reduction of the moment of inertia of a solid sample, with respect
to its classical value. This effect is commonly referred to as Non-Classical Rotational
Inertia, or NCRI. Consider a cylindrical vessel (cell), filled with a given substance, and let the
cell be connected to a torsion rod, allowing it to execute small harmonic oscillations about its axis, as shown in Fig. \ref{fI1}. This experimental set-up is typically referred to as {\it Torsional Oscillator} (TO).
\\ \indent
The measurable resonant period of oscillations is proportional to the square
root of the moment of inertia of the system, which of course includes a
contribution from the vessel itself, as well as from the substance enclosed in it.

 If below a given temperature, the resonant period shows a monotonic, continuous decrease, one way to interpret this is to assume superfluid
 mass {\it decoupling} from the rotation of part of the sample inside the vessel. In other words, a fraction thereof (normal) keeps oscillating together with the apparatus, while the remaining (superfluid) part remains at rest, therefore not contributing to the rotational moment of inertia. Formally,
\begin{equation}
I(T)=I_{\rm class}(1-\rho_s(T)/\bar \rho)\;.
\label{NCRI}
\end{equation}
where
$I_{class}$ is the classical moment of inertia, regarding the whole sample as normal. Thus, a measurement of the resonant period is a measurement of the superfluid fraction of the system, as long as $I_{\rm class}$ and
the elastic properties of the entire apparatus remain constant
as the temperature is lowered.
\\ \indent
The NCRI  technique is particularly well-suited to probe possible superfluidity of a solid, as it does not require any extra circuitry to produce relative flow between the normal and superfluid components.
However, the NCRI-TO experiment is not without some ambiguity, in part precisely because it is  indirect, i.e., it does not involve any actual measurement of mass flow
out of the system. Indeed, a number of different physical mechanisms have been proposed, not involving a superfluid transition of  the helium crystal,  that could conceivably cause a change in the oscillation period, thereby giving rise to a spurious NCRI signal \cite{clarkh2}. An example of such a mechanism is given by a possible structural transformation taking place in the crystal at low temperature. For these reasons, careful, concurrent control measurements on similar, non-superfluid systems\footnote{For example, in the case of solid $^4$He an obvious control experiment consists of performing the same measurement on solid $^3$He, a Fermi system not expected to undergo a supersolid transition -- certainly not at the same temperature as $^4$He.} are required, in order to ascertain by comparison whether what
is being observed is indeed a genuine superfluid transition.
\\ \indent
There are a number of puzzling observations, connected with the onset of NCRI, which have so far eluded a satisfactory explanation. We come back to this point below, when discussing experimental evidence of  supersolidity of $^4$He that is available at this time, NCRI measurements constituting the bulk thereof. We shall propose a possible theoretical interpretation of (at least some of)  the phenomenology in that context, in Sec. \ref{sec:he4}.

\subsubsection{Mass transport through a supersolid}
\begin{figure}[t]
\includegraphics[ width=3.0in]{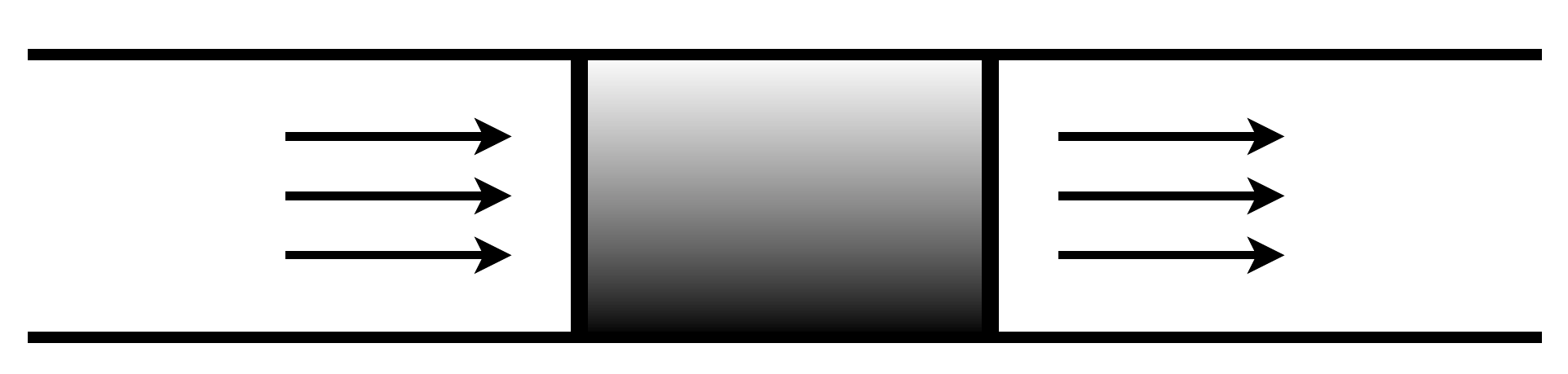}
\caption{{\it Color online}. Schematic of an experiment measuring mass flow through a supersolid. Superfluid liquid flows  across  the central section filled with supersolid, of the same substance. The setup as shown can work only at the melting pressure.
In the case of $^4$He, modifications involving Vycor segments and temperature gradients allow one to extended measurements to higher pressures \cite{Ray}.
}
\label{fI2}
\end{figure}

A different way of detecting supersolid behavior is based on flow,
not {\it of} the supersolid, but {\it through} the supersolid.
Consider the experimental set-up schematically shown in Fig. \ref{fI2}. One wishes to establish flow of a fluid through a pipe, a section of which is filled with solid of the same substance. The most obvious example, one with which we are all familiar, is that of a water duct. During a cold Winter month, water can freeze in a  section of the duct. As we know from experience,  water flow (for example, in the direction shown by the arrows in Fig. \ref {fI2}), ceases as soon as ice (solid water) clogs the duct. There is, of course, nothing peculiar about water in this regard; the same observation will be made if any other substance fills the pipe, as long as freezing takes place as shown. Indeed, not even {\it superfluid} liquid helium may flow through the same section of the pipe, if filled with {\it normal} solid helium.
\\ \indent
On the other hand, {\it some} flow will take place through  {\it supersolid} {helium}.
In other words,  a superfluid liquid will flow through a supersolid
{\it made of the same elementary constituents}.
The above experiment allows one to gain intuitive, vivid understanding of the somewhat cryptic idea of ODLRO, as expressed by condition (\ref{odlro}); for,  one may think of helium atoms in the superfluid phase, flowing toward the frozen blockage (left part of Fig. \ref{fI2}),  essentially ``re-emerging" on the right side of it, through a process of long exchanges with (indistinguishable) atoms in the supersolid.   Thus, ``flow" through a supersolid necessarily must involve identical particles, as it is underlain by quantum exchanges. An impurity, namely a particle distinguishable from those that make the supersolid, will not move through  it any more than through a regular crystal.
In this respect, a supersolid acts very differently from a superfluid,
which allows a foreign particle to move through it with very little drag
(no drag at all at $T=0$).
\\ \indent
Experiments such as the one described above, have been performed on solid helium only relatively recently \cite{Ray}. They are also not without
drawbacks, e.g., the difficulty of inferring a reliable value for the superfluid fraction. We shall describe the main results of these experiments, and the progress that they have afforded on the understanding of supersolidity in $^4$He, in Sec. \ref {sec:he4}.
\subsubsection{ Momentum Distribution}
The momentum distribution $\tilde n({\bf k})$ (Eq. \ref{nofkdef}), can in principle expose in full the supersolid character of a system. For a  crystal of ``Boltzmannons", namely {\it distinguishable quantum particles}, $\tilde n({\bf k})$ is a Gaussian, whose semi-width is determined by zero-point motion.
Particle indistinguishability and quantum statistics cause measurable deviations of the momentum distribution from the Gaussian behavior.
Specifically, the one-body density matrix $n({\bf r})$, which is the Fourier transform of $\tilde n({\bf k})$, decays exponentially in a normal (i.e., non-supersolid) Bose crystal, conferring to $\tilde n({\bf k})$ a much slower decay with $k$ than the Gaussian one \cite{bon09}.
\\ \indent
In the presence of ODLRO, $n(r)$ settles asymptotically on the finite value $n_\circ$ at large $r$. In the case of a supersolid, however, it must also take on an oscillatory behavior, reflecting lattice periodicity. This is because, in a solid, long many-particle exchanges which underlie ODLRO involve particles
situated preferentially at lattice sites. That means that $n({\bf k})$ in a supersolid will feature satellite peaks at reciprocal lattice vectors.
\\ \indent
Thus, in principle a measurement of $\tilde n({\bf k)}$ could provide the most direct assessment of whether a system is supersolid or not, as both types of order are probed simultaneously. In practice, however, the feasibility of such a determination depends on the strength of the signal, i.e., $n_\circ$,
as well as on the inherent sensitivity of the measuring apparatus.
\\ \indent
For example, the momentum distribution of solid helium can be studied experimentally by deep inelastic neutron scattering. The resolution of the most advanced experimental facilities available nowadays, as well as the uncertainty associated with the analysis of the raw data and the expected, sheer feebleness of the supersolid signal, make it all but out of question to address the issue of supersolidity of $^4$He at the present time. Moreover, there exist more robust techniques (e.g., x-ray diffraction) that allow one to establish unambiguously whether a condensed matter system of interest has crystal order; it therefore makes sense to carry out separate measurements, aimed at probing individually superfluidity and density LRO. Because in the presence of interactions, the superfluid fraction is usually greater in magnitude than the condensate fraction, there is generally a preference among experimentalists for measuring the former.
\\ \indent
In a different physical contexts, however, the measurement of $\tilde n({\bf k})$ is inarguably the most efficient way of establishing the supersolid
nature of an underlying many-particle systems. This is typically the
case for assemblies of cold atoms, whose  momentum distribution
lends itself to direct imaging, by means of time-of-flight measurements \cite{bloch}.
\\ \indent
\subsection{Theoretical studies}
From the theoretical standpoint, the investigation has proceeded along several fronts. This Colloquium is {\it not} aimed at providing an overview and/or critical comparison of {\it all} the different theories. For example, we shall not review nor discuss any of the scenarios which attribute the observed NCRI  to  effects that are mechanical in origin, and/or in any case not associated  to any superfluid response of {\it solid} $^4$He \cite{yoo,Reppy10,blabla2}. Likewise, we shall not review the ``vortex liquid" theory\footnote{It is worth mentioning that, while the ``vortex liquid" theory, which assumes a superfluid ground state for an ideal $^4$He crystal, is in direct contradiction with existing numerical evidence, it contains some of the same mesoscopic phenomenology 
present in the so-called Shevchenko state \cite{Shevchenko87}, discussed below.}  \cite{contento,adesso}. 
All of these theoretical proposals are the subject of intense current debate, and interested readers should check the pertinent references.  Instead,  we shall  restrict here our discussion to scenarios that attempt to connect experimental observations to  {\it first principle} calculations, carried out on increasingly realistic models of a helium crystal.
\subsubsection{First principle calculations}
The expression ``first principle", in this Colloquium,  means that one starts from a quantum-mechanical many-body
Hamiltonian $\hat H$ for a system of $N$ identical particles, namely:
\beq\label{ham}
\hat H  = -\frac{\hbar^2}{2m}\sum_i\nabla^2_i + V({\bf r}_1,{\bf r}_2,...{\bf r}_N)
\eeq
where $m$ is the mass of each particle in the system, and where $V$ is the potential energy of interaction. In all cases of interest here, $V$ can be expressed as the sum of terms involving pairs of particles, i.e.,
\beq
V({\bf r}_1,{\bf r}_2,...{\bf r}_N) = \sum_{i <j} v({\bf r}_i, {\bf r}_j)
\eeq
where the function $v({\bf r}_i,{\bf r}_j)$ is known either from accurate {\it ab initio} quantum chemistry calculations (e.g., the interaction between two helium atoms by \textcite{aziz79}), or, in the case of cold atoms, can be as simple and fundamental as the potential of interaction between two electric dipoles.
\\ \indent
The calculation of cogent physical quantities, characterizing {\it equilibrium} thermodynamic phases of many-body systems modeled as in Eq. \ref{ham}, can be carried out to a remarkable degree of accuracy by means of numerical simulations, based on Quantum Monte Carlo techniques (see, for instance, Ref. \textcite{worm}). This computational methodology is fairly mature, and thoroughly described in the literature; we therefore refer interested readers to the original references. What is important to stress here, is that these methods allow one to obtain direct, unbiased numerical estimates of {\it all} the quantities that are relevant in the study of the supersolid phase of matter, namely the one-body density matrix $n(r)$, the superfluid fraction $\rho_S(T)$ and the static structure factor (standard finite-size scaling procedures allow one to extrapolate results to the thermodynamic limit).
Essentially all of the theoretical results presented and discussed in the following sections, are obtained by means of Quantum Monte Carlo simulations. 
\subsubsection{Lattice Supersolids}
A different approach, also affording fundamental insight into the nature of
the supersolid phase, the transition to a supersolid from a normal solid or from a superfluid, and the
role of point defects, makes use of much simpler models than Eq. \ref{ham},
wherein particles are confined to moving on {\it discrete lattices}.
In this case, spontaneous breaking of translation invariance is defined with reference to the discrete translation symmetry of the Hamiltonian.
The simplicity of the models utilized, and the possibility of carrying out very accurate numerical calculations, make it possible  to address a number of theoretical issues relevant to the problem of supersolidity, in some cases offering a valuable guide to the experimental investigation. Indeed, some of the earliest microscopic theoretical investigations aimed at assessing the possible existence of a supersolid phase of $^4$He, made use of lattice Hamiltonians \cite{matsuda,liufisher}.
Moreover,  while for a long time lattice models were regarded as little more than a useful theoretical device,  impressive advances in optical lattice technology render it now possible to synthesize and investigate in the laboratory artificial many-body systems,
very accurately realizing  the physics embedded in those models.
Therefore, one may conceivably be able to carry out a direct comparison
of theory and experiment, enjoying a degree of accuracy not attainable in solid helium experiments. 
\\ \indent
There exist, however, fundamental differences between continuous
space and lattice supersolids, which inevitably limit the
understanding that any one of them can afford of the physics of a system in continuous space.
A lattice ``crystal" is one in which basis vectors are linear combinations of those of the underlying lattice.
Such a solid period cannot be varied continuously, i.e.,
a lattice crystal is by definition incompressible.
Consequently,  supersolid phases
can be obtained in a rather trivial way, simply by doping the insulating crystal
with holes or particles (provided that no phase separation occurs of the doped
system into a superfluid and an undoped crystal \cite{batrouni}).
While continuous space crystals can lower their energy by eliminating
vacancies and interstitials, using complete atomic layers and adjusting the
lattice constant accordingly, an analogous mechanism is forbidden if crystalline
sites must remain pinned to an underlying space lattice. Consequently,
in lattice crystals point defects can be introduced ``by hand";
their presence is not necessarily of ``zero-point'' origin,
and their number is conserved.
The second crucial difference between continuous and pinned lattice solids,
is the absence in the latter of long-range elastic forces between point defects.
Thus far, all lattice supersolids studied in the literature (in fact,
we are not aware of a single counter-example)
belong to the category of doped insulators, which is fundamentally
different from the zero-point defect picture
relevant for continuous space systems.
Heretofore, we restrict our discussion to continuous-space
supersolids, making use of lattice systems occasionally, for
illustration purposes only (e.g., in Sec. \ref{ssec:defects}). Readers interested in lattice supersolids are
referred to the vast literature on this subject, of which  \textcite{batrouni,hebert,ycchen,dang,pinaki,boninsegni,melko,wessel,damle,kolya}
are just a representative sample.
\\ \indent
In the following two sections, we shall {\it a}) review
the present status of the  theoretical and experimental research on supersolidity in
$^4$He, and {\it b}) discuss theoretical predictions for supersolid behavior in systems of cold atoms.

\section{Supersolidity in \he4}
\label{sec:he4}

The experimental search for the supersolid phase of matter, has focused
over the last four decades almost exclusively on the most quantum
solid in Nature, whose elementary constituents obey Bose statistics, namely \he4. Helium, unlike any other naturally occurring substance, remains a
liquid under the pressure of its own vapor, all the way down
to zero temperature. It solidifies under some moderate pressure,
(approximately 25 atm), forming a hexagonal close-packed (hcp) crystal. It is worth noticing upfront two  aspects, that complicate the unambiguous detection of supersolid behavior in a helium crystal. The first is the resilience of superfluidity in the liquid phase, which is predicted to remain superfluid in the low temperature limit at arbitrarily high density, i.e., in the metastable overpressurized liquid \cite{moroni}. The second is the presence in an ordinary helium crystal of  highly mobile, extended defects. Thus,  any manifestations of superfluid behavior of a crystal of helium may be inherently difficult to disentangle from the  response of  persistent pockets of liquid.
\\ \indent
Initial attempts to detect the possible supersolid character of solid $^4$He yielded negative outcomes \cite{Meisel}. However, some of those experiments
were based on the misconception that a supersolid would act {\it exactly} like a superfluid, e.g., eliminate any pressure differences between two cells connected by a
narrow channel (``superleak"), or allow motion of macroscopic external
objects through the crystal. As discussed in Sec. \ref{exp},
a true supersolid will only allow  dissipation-free flow of its own constituent
particles; with respect to any other object, its behavior
is no different from that of a regular solid.
For example, a steel ball will not sink under gravity to the bottom
of a vessel filled with supersolid; nor will a pressure gradient disappear, either established within the supersolid sample, or
between two connected chambers, as long as the lattice structure itself remains kinematically
stable. Of course, pressure differences will result in the redistribution of
the superfluid component, so as to ensure constancy of  chemical potential
throughout the system, for a given pressure
field $P({\mathbf r})$. However, if the superfluid fraction is small such a redistribution will
only result in tiny corrections to the original pressure gradients,
$\nabla P({\mathbf r}) \to \nabla P({\mathbf r}) (1-\rho_S/\bar\rho)$.
This is, in essence, an analog of the fountain effect,
but with an additional ingredient
that the normal component is immobile even in the bulk (excluding plastic flow from the picture).
\\ \indent
At the time of this writing, there is no coherent understanding of
the observed period shift in torsion oscillator experiments.
There exist a number of experimental observations which render
problematic a straightforward interpretation of
Kim and Chan's findings in terms of supersolid $^4$He.
For example, it is found that the period shift occurs in concomitance with a stiffening of the crystal \cite{Beamish,dsb};
that it is sensitive to the resonant frequency of  the apparatus;
that it lacks the characteristic critical behavior
at the alleged superfluid transition temperature; that it is present in TO experiments in which blockages are inserted, which should radically
affect the superfluid response; that it is affected by the presence of small concentrations of $^3$He impurities, as well as by DC rotation; and that it can be even incorrectly
identified, due to the delicacy  of high-temperature ``background'' subtraction \cite{Reppy10}.
\\ \indent
We shall thus focus in this paper, on what we believe to be reliable theoretical results for
hcp crystals of $^4$He,  and on the possible scenario of interconnected
superfluid network in structurally disordered crystals, as a plausible
explanation for at least some of the present experimental evidence. This scenario
is also supported by recent mass transport experiments by Ray and
Hallock \cite{Ray}.

\subsection{Perfect hcp crystals of $^4$He}
\label{ssec:he4}

\begin{figure}[tbp]
\centering
\includegraphics[ width=3.0in]{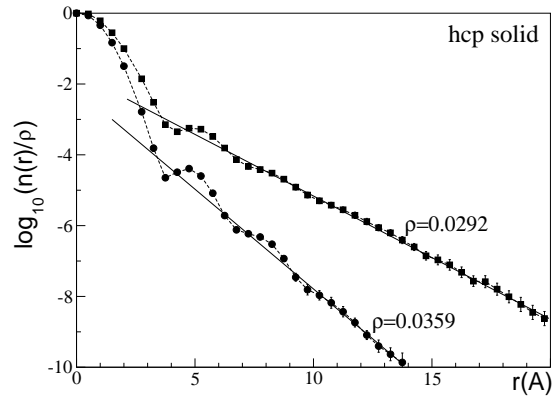}
\caption{One-body density matrix $n(r)$, computed by Monte Carlo simulation, of solid (hcp) \he4 near (density $\bar \rho =  0.0292$ \Am3)
and away from (density $\bar\rho =0.0359$ \Am3) the melting curve at a low temperature
T = 0.2 K. Results obtained at lower temperature (as low as 0.1 K) are not distinguishable from those shown here, within the statistical uncertaities of the calculation. The exponential decay of $n(r)$ signals no off-diagonal long-range order (i.e., Bose-Einstein Condensation) in the system. Reprinted from Ref. \cite{B1}.}
\label{fig:1}
\end{figure}
Fig.~\ref{fig:1} shows the one-particle density matrix $n(r)$  of an ideal hcp $^4$He
crystal at low temperature T = 0.2 K, computed by
Monte Carlo simulations at two densities, i.e., near the melting
curve and at relatively high pressure. By ``ideal", we mean a {\it commensurate}
crystal, i.e., one with exactly two atoms per unit cell; no point defects such
as vacancies or interstitials, nor extended defects such as dislocations, disclinations, or grain boundaries, are included.
These simulations are based on a microscopic model of helium (Eq. \ref{ham}), making use of the accepted Aziz pair potential \cite{aziz79}.
\\ \indent
The results for $n(r)$ shown above
are independent of temperature (below 2 K, down to at least 0.1 K), as well as on the number of particles
in the crystal, at least for distances $r < L/2$, where $L$ is the linear system size \cite{B1}. The observed temperature-independent exponential
decay constitutes robust evidence
that the system is insulating in the ground state (i.e., non-superfluid). Such an exponential decay has been observed in {\it every single calculation} of the one-particle density matrix in {\it commensurate} solid $^4$He, based on {\it unbiased} Quantum Monte Carlo methods. This includes not only finite-temperature calculations \cite{C1,cazzoni}, but also those based on ground state projection \cite{galli}.
It is important to note that considering a commensurate number of particles, cannot
have any effect on the properties of the supersolid phase, in which the number of particles
is a {\it continuous} function of the chemical potential.
Insulating crystals, on the contrary, can change particle number
in response to small changes in $\mu$, at constant volume, only by
adding/removing complete atomic layers. In other words they (i)
always stay commensurate, and (ii) have zero isochoric
compressibility since the process of adding/removing atomic layers is
kinematically frozen in an ordered state.
\\ \indent
A comment is in order at this point, since in this Colloquium we only discuss  numerical evidence such as the one displayed in Fig. \ref{fig:1}. 
In our opinion, these results qualify as  ``proof" (though not 
in a strict mathematical sense) that a perfect $^4$He crystal does not possess off-diagonal long-range order. Most numerical studies have well-known limitations, chiefly the finite size of the system that can be simulated using modern computing facilities. There exist, however, well-defined ways (i.e., finite-size scaling analysis) to deal with such limitations, and essentially remove them by assessing their effects on the physical results obtained, in the thermodynamic limit. It seems objectively difficult to dismiss outright the results 
of Quantum Monte Carlo simulations, in light of the quantitative accuracy with which they have been shown to reproduce many measurable properties of helium and other quantum solids (see, for instance, \textcite{cepe}). Though the validity of the prediction of absence of ODLRO based on simulations has been called into question \cite{madai}, it seems fair to state that, given the difficulty of obtaining a formal analytical proof,  numerical results such as those quoted here presently constitute the strongest foundation, on which as unbiased as possible a statement can be made, without making any {\it a priori} assumption on the physics of the system.
\\ \indent
The exponential decay of the density matrix is consistent with finite
energy gaps for creating vacancies and interstitials. These are readily
inferred from the exponential decay of the single-particle imaginary-time (Matsubara)
Green function,
\beq
G({\mathbf p},\tau ) = - \langle {\cal T} [\hat \psi^{\;}({\mathbf p},\tau )
\hat\psi^{\dag}({\mathbf p}, 0 )]\rangle
\eeq
where ${\cal T}$ is the imaginary time ordering operator and $\hat\psi({\bf r},\tau)$ is a time-dependent
field (see, for instance, \textcite{fetter}). The above quantity was also computed by Monte Carlo, giving at the
melting density $\Delta_V \approx$ 13 K and $\Delta_I \approx$ 23 K
for vacancy and interstitial formation gaps, respectively \cite{B2}.
The insulating gap $\Delta_I+\Delta_V$  was found to increase with pressure,
as expected. \\ \indent
The Andreev-Lifshitz supersolid scenario is based
on gapless vacancies or interstitials, with repulsive interactions.
The large values of  $\Delta_V$ and $\Delta_I$, by themselves, all but rule it out.  Moreover,
effective interactions among vacancies\footnote{The
same results obtained for vacancies are found for interstitials too. Henceforth, we shall restrict
our discussion to vacancies.} are found to be strongly attractive \cite{B2},
causing the formation of bound states of two, three, etc. vacancies.
Attractive interactions among vacancies
will lead to their collapse, the resulting final state
depending on vacancy concentration and/or system density. For initial densities
below freezing, i.e., $\bar\rho<\rho_f \approx 0.0259$ \Am3, the final state will be
all liquid; for initial density between freezing and melting,  i.e.,
$\rho_f<\bar\rho<\rho_m \approx 0.0287$ \Am3, one will observe phase separation
into liquid and solid domains; finally, for densities $\bar\rho>\bar\rho_m$, the final state
will contain dislocation loops produced by coalescing vacancies.
Even if out-of-equilibrium vacancies are introduced into the
crystal at some small finite density, they will not be able to form a
long-lived metastable condensate, similar to that
which exists in cold atomic gases.
The crucial difference between the two systems, lies in the presence
of the solid matrix in helium, playing the role of ``third body'' required to satisfy
energy and momentum conservation, in recombination processes leading to
clustering of vacancies.

The above results constitute strong evidence that a perfect hcp crystal of $^4$He
is an insulating solid. By ``perfect", we mean a single crystal, without extended
structural defects such as dislocations, disclinations, and grain boundaries.
In turn, this suggests that, {\it if} a hcp crystal does have the ability of supporting dissipation-less flow of
its own atoms, this must necessarily take place along extended structural defects.
\subsection{Superfluidity of crystalline defects}
\label{ssec:defects}

Consider a many-body system in which the motion of particles is confined to one or two spatial
dimensions. For example, one could imagine an assembly of cold atoms with tight optical confinement
in one or two spatial directions, or a helium film adsorbed on a substrate, or a superconducting
wire. In these cases, the rationale for ignoring the dimensions that are ``blocked", is that the systems
are either {\it in vacuo}, or embedded in insulating materials made of different atoms or molecules.
When it comes to extended, structural defects in a crystal, however, which can be regarded as many-body
systems in reduced dimensions, superfluid properties are inseparable from those
of the insulating bulk system. This is because the same particles
which form an insulating state in the bulk, are responsible for
superfluidity in the lower-dimensional defect structure.
\\
Realistic systems are rarely perfect on the macroscopic scale;
domain walls form naturally in the out-of-equilibrium cooling process
across the transition temperature.
One may wonder whether transport properties of extended defects in quantum solids
are fundamentally different from the  matrix in which they are embedded.
An insulating state is typically the result of competition between kinetic and
increasing potential energy terms. One can imagine cases when in some finite vicinity
of the liquid-to-solid transition in the bulk system, even small changes in the
arrangement of atoms change the balance in favor of particle delocalization.
For example, potential barriers for moving atoms along the structural defect may
be substantially lower than in the bulk.

In order to gain some understanding of this physical mechanism, consider a model of lattice hard core bosons,
with a repulsive interaction $V$ between two particles occupying adjacent lattice sites. Let $-t$
be the value of the matrix element of the Hamiltonian corresponding to the hopping of a particle to a
nearest-neighboring site. In a square lattice geometry and at half filling (i.e., at a density corresponding
to one particle every two sites), a ``checkerboard" lattice will form when
$V >V_c= 2t$.

\begin{figure}[tbp]
\includegraphics[ width=3.3in]{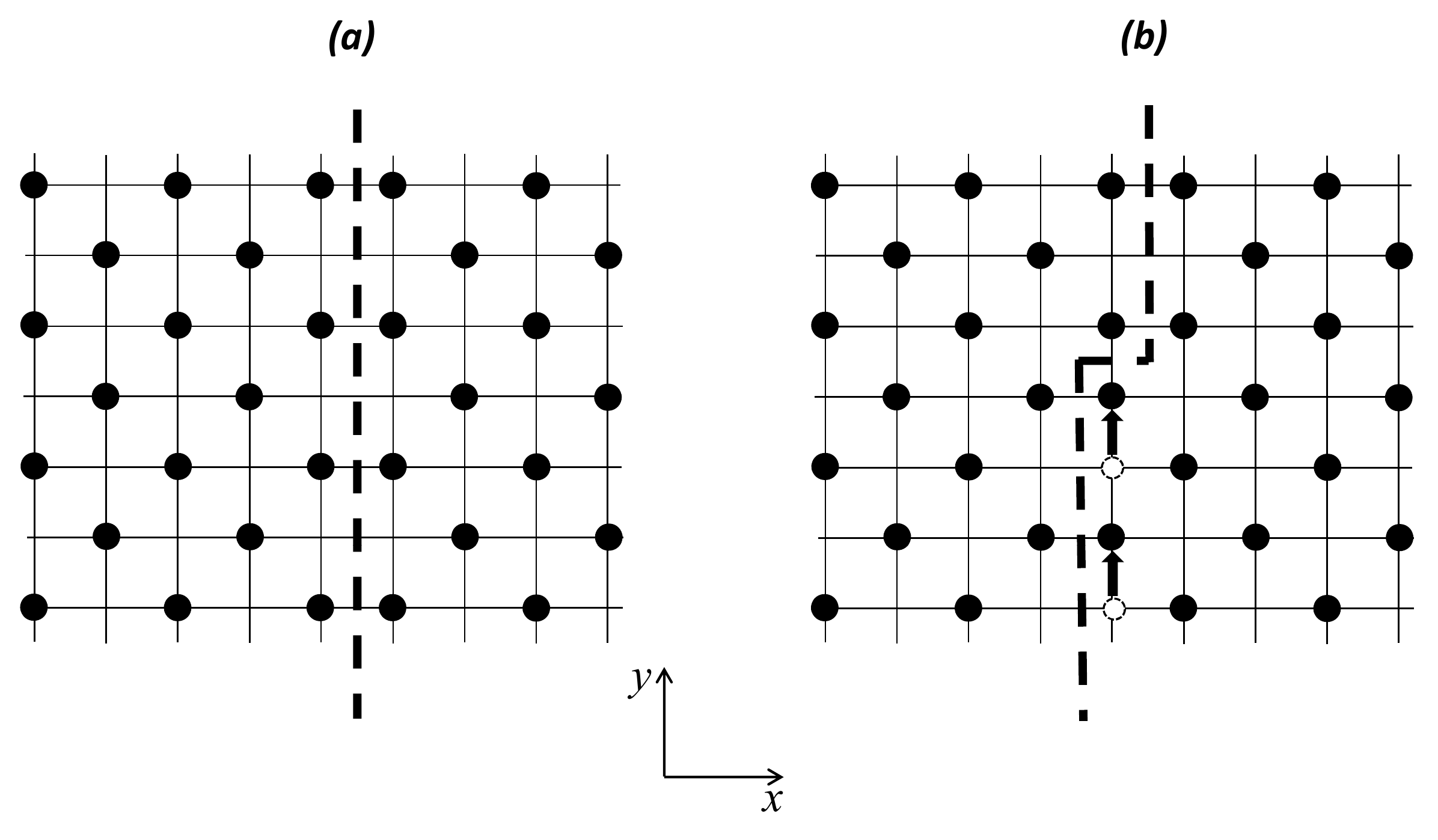}
\caption{ (a) Domain wall in the checkerboard solid. (b) When an atom is moved along the
wall by one atomic distance, the wall position shifts in the perpendicular direction while
the kink on the wall moves along the wall by two atomic distances.
}
\label{fig:2}
\end{figure}

Fig. ~\ref{fig:2} shows a domain wall in a checkerboard crystal \cite{GB1,GB2}. Such a defect
automatically appears in this lattice model, if periodic boundary conditions are adopted and
an odd number of columns is assumed, as in Fig. ~\ref{fig:2}.
When an atom in the configuration shown is shifted to a nearest neighbor position
in the bulk, the potential energy goes up by $3V$. Shifting an atom by one
atomic distance along the wall will increase potential energy only by $V$.
Clearly, these considerations crucially depend on both the bulk and the defect
structure, as well as the type of the motion considered.

It is also easy to see from this example an intricate connection between
the motion of atoms along the defect and the motion of the defect itself
in the bulk. If we take the classical picture shown in
Fig.~\ref{fig:2}(a) and shift up by one atomic distance up all atoms
nearest and to the left of the dashed line then the wall position will
shift to the left in the transverse direction, see Fig.~\ref{fig:2}(b).
As we will see shortly, this brings about interesting physics
relating superfluidity, roughening, and quantum plasticity.

Finally, superfluidity along structural defects which form
an interconnected three dimensional network may be considered as yet another
scenario for obtaining the supersolid state. The corresponding state
clearly breaks both translation and U(1) symmetries even though it is not in
thermal equilibrium. Still, since extended defects such as domain walls,
grain boundaries, and dislocations are protected by the lattice topology
the lifetime of their structural relaxation may easily exceed all relevant
experimental time scales.

\subsection{Domain walls}

Most transitions into the solid phase are first-order in nature, and, in general, there
is no reason why domain walls or grain boundaries in solids should have
superfluid properties. It is the right combination of light atomic mass,
weak inter-atomic forces, and degree of structural frustration, which allows
atoms to establish long-range coherence at the defect core.
For the domain wall shown in Fig.~\ref{fig:2}(a)
one can argue, however, that at least for $V-V_c \ll t$,
superflow of atoms along the domain wall will surely take place \cite{GB1}.
This is because for a system of hard-core bosons with nearest-neighbor
interactions, the superfluid-solid transition is marginally
first-order. That is, the critical point itself has the same extended SU(2)-symmetry as the  isotropic Heisenberg model.

As the interparticle interaction is increased beyond a threshold value
$V_W$ ($V_W = 3.57(3) t$ in $d=2$ and
$V_W = 2.683(3) t$ in $d=3$ \cite{GB2}) one observes
the superfluid-to-insulator transition (SF-I) within the wall,
with the concomitant disappearance of gapless sound excitations.
In the limit of large $V/t$,
the classical picture of Ising-type domain wall shown in
Fig.~\ref{fig:2}(a) is recovered. For two-dimensional walls in 3D solids,
no obvious reason seems to exist for relating superfluidity to structural transformations
of walls and grain boundaries, at the SF-I transition point.

The same considerations apply to any domain wall, twin, or generic grain
boundary in solids, with an important {\it caveat}---most solids form through a
strong first-order transition, and the resulting crystalline structures are typically
very efficient at localizing atoms.
For example, even in the highly quantal helium crystals at the melting curve,
the smallest exchange cycles are suppressed relative to the liquid phase, by
several orders of magnitude (despite the large zero-point motion of individual
atoms). As a figure of merit, one may mention the energy scale for the tunneling
matrix element of $^3$He impurity in $^4$He solid, $J_{34} \sim 10^{-4}$ K
\cite{Guyer,Mullin}. It is more than five orders of magnitude smaller than the Debye
frequency characterizing large zero-point motion of atoms, leading to
an effective mass of a $^3$He impurity which is about four orders of
magnitude heavier than a helium atom in vacuum. In other words, structural
order emerging from first order transitions in most solids is so strong, that
grain boundaries and other extended defects are likely to end up on the
insulating side. Helium is a unique element in this regard, since it is the
only one remaining in the liquid phase at saturated vapor pressure. Though experimental
evidence for superfluidity of grain boundaries in $^4$He is rather incomplete
\cite{BalibarR}, Quantum Monte Carlo simulations do indicate that this possibility
is real \cite{GB3}.

One-dimensional walls in 2D solids allow for
an interesting connection between superfluidity and roughening, at least in some cases. Naively,
one might think of these two phenomena as being completely unrelated,
as they pertain to
different types of motion (i.e., along and perpendicular to the wall direction). However, in some cases they may have exactly the same microscopic origin
when neither is possible without the other. As shown
in Fig.~\ref{fig:2}(b), the simplest, and energetically least expensive deformation
which results in the displacement of the wall in the ${x}$-direction, consists of
creating a ``kink and anti-kink'' pair, by shifting atoms along the wall.
Clearly, from such a deformation mass transfer along the wall inevitably ensues.
This suggests the intriguing
possibility that superfluidity may occur in concomitance with a roughening transition.
By ``rough" we imply a line with large
mean square shape fluctuations $(\Delta x)^2$  (actually diverging in the thermodynamic limit).
More quantitatively,
if the instantaneous wall coordinate is described by $x(y,t)$ then
\begin{eqnarray}\nonumber
(\Delta x)^2 = \langle L^{-1}\int_0^{L_y} dy\ [x(y,t)- \bar{x} ]^2 \rangle \;, \\ {\rm with}\ \
\;\;\; \bar{x} = L^{-1}\int_0^{L_y} dy\ x(y,t) \;,
\label{DW2}
\end{eqnarray}
where the average is taken over the statistical ensemble. The line is in the so-called
``smooth'' state, wherein $(\Delta x)^2$ is system size independent.

Interestingly enough, roughening helps to stabilize the superfluid phase.
One might think, by looking at the classical picture of a straight wall shown in Fig.~\ref{fig:2}(a),
that checkerboard order in the bulk {\it de facto} doubles the size of the unit cell, and
therefore the effective occupation number
in the wall is integer. Notice, however, that the bulk potential is
shifted by half a period when the wall is displaced by one lattice spacing in the
${x}$-direction, meaning that for rough walls
the average potential still retains the translation symmetry of the original lattice. Thus,
the effective filling factor remains half-integer.
This significantly increases the parameter range within which superfluidity survives.
Numerically, one indeed finds that superfluid and roughening transitions
coincide \cite{GB2}.

\subsection{Screw and edge dislocations in $^4$He}

There are infinitely many distinct dislocation lines in the solid
structure. It was conjectured by S. Shevchenko \cite{Shevchenko87} that
edge dislocations in $^4$He may be superfluid, but in the absence of
first-principle simulations it is impossible to predict which edge dislocations,
if any, have that property. It is found by Monte Carlo simulations, however, that most
edge dislocations are in fact insulating, their core splitting into two partials.
A split-core configuration substantially reduces frustration
in the atomic arrangement, and locks atoms in an insulating state \cite{Hestress}.
Instead, the most basic screw dislocation, oriented along the ${c}$-axis
of the hcp structure, is found by simulations to support strong superfluid flow
along the dislocation line \cite{Hescrew}.
At the melting point, the effective one-dimensional superfluid
density is about $\rho_S^{1d} \approx$ 1 \AA$^{-1}$ corresponding
to a ``superfluid tube'' at the core of diameter $\approx$ 6 \AA.
The corresponding Luttinger liquid parameter, $K \approx 4.9(5)$, is large  enough
to ensure that superfluidity of screw dislocations be stable against weak
disorder or a commensurate potential.
\\ \indent
Since screw dislocations are very common in crystals,  and often facilitate
crystal growth, it is not unreasonable to expect that a network of superfluid ``pipes'' penetrating the insulating bulk
may exist in real samples of solid
$^4$He, capable of conducting mass currents of $^4$He atoms,
without measurable dissipation.
An obvious issue that any theoretical scenario based on a
network of dislocations must address, is whether
different types of dislocations with comparable superfluid
properties may exist, in order for a fully connected
three-dimensional network to be established.
As it turns out, edge dislocations with Burgers vector along the hcp ${c}$-axis
and the dislocation core along the ${x}$-axis in the basal plane, also
display in simulations a finite, albeit weak, superfluid response \cite{Heedge}.
In this case, the core is split into two partials
separated by extremely large distance (possibly as large as 150-200 \AA)
and the superfluid signal is ``marginal'', i.e., the corresponding Luttinger
parameter is close to the critical value $K=2$
(with a large uncertainty) for which a commensurate potential becomes a relevant
perturbation. The large separation between the two half-cores is explained by the
very small (practically unmeasurable) surface energy of the stacking fault defect in the
hcp structure, leading to
a very weak linear potential confining partials.
The unusual split-core structure may have interesting consequences for how edge
dislocations move, anneal, and are pinned in quantum crystals; these issues still
need to be fully explored, at the time of this writing.
\\ \indent
Superfluidity in the cores of edge dislocations has important consequences
in terms of  how solids react to chemical potential differences, leading to a
novel physical behavior, to which we refer as ``quantum metallurgy'' 
(though helium is obviously {\it not} a metal)\footnote{We attribute this term to A. Dorsey}. At
temperatures well below $\Delta_V$, i.e., in the absence of activated vacancies,
non-superfluid edge dislocations cannot move in the climb direction
perpendicular to their axis and Burgers vector. For, such a climb would correspond
to the growth of an atomic layer which, in turn, requires mass transfer.
Since activated vacancies is the only mechanism of atomic mass
transport in insulating solids, one concludes that standard edge dislocations
ought not climb. This, in particular, means that insulating crystals
are {\it  isochorically incompressible},
$\chi\equiv (d\bar\rho/d\mu)_V \approx 0$; in other words, the density $\bar\rho$ of
the crystal does not change atom by atom 
\footnote{In the absence of vacancies and interstitials,
the density of a crystal can react dynamically to small changes in the
chemical potential, $\delta \mu$, only by creating or removing crystalline layers.
This requires nucleation times exponentially large in $|\delta \mu|^{-1}$.} in response to infinitesimal, quasi-static changes in $\mu$.
\\ \indent
If cores of edge dislocations in the network are superfluid, then efficient transport of
atoms  throughout the sample is possible, and superfluid edge dislocations
may to climb (or, {\it superclimb}).
Correspondingly, the crystal density ought to react continuously to small changes
in chemical potential, and an anomalous
isochoric compressibility of a material which is insulating in the absence
of structural defects, should be observable. This description precisely matches
the observation of  Ray and Hallock, who carried out an experiment wherein $^4$He atoms
(superfluid liquid) were fed into the crystal through implanted Vycor rods \cite{Ray,Ray2}.
In such a setup, the chemical potential $\mu$ is the physical quantity relevant
to the external perturbation applied to the crystal. Also, the increase in the crystal
density was correlated with the ability of solid samples to conduct atoms
between the Vycor rods.

\subsection{Superglass}
As mentioned above, extended structural defects are metastable excited states
of the system, which are protected by topology of the crystal. One may wonder to what
a degree this picture may continue to hold,  as the number of defects increases,
to the point where the original solid structure may barely be recognizable; to take this
one step further, one may even consider a metastable, {\it glassy} phase arising from the application
of an instantaneous compression to a superfluid.\\
Such a highly disordered state may be short-lived, and quickly relax to a polycrystal
with a clearly identifiable mesoscopic solid structure (here, ``quickly" is defined with
respect to  relevant experimental time scale). However, this need not be necessarily
the outcome, for a glass produced from a sudden compression, such as described above,
may actually be stable on the relevant experimental (or {\it any} meaningful)
time scale\footnote{Many amorphous solids are among the
strongest in Nature, with respect to mechanical shear or deformation.}.

According to the conventional wisdom, at low temperature glasses are insulators.
For, if atoms were
allowed to move around, then the system would quickly discover a
(poly)crystalline arrangement lower in energy. However, the underlying
assumption  that {\it any} kind of atomic delocalization will eventually lead
to formation of polycrystalline seeds, is not justified. There is actually no fundamental
reason to assume that jamming of structural relaxation should be incompatible
with superflow.
In fact, one can imagine a system where these two types of atomic displacements
occur on vastly different time scales, in such a way that the glass may remain structurally
stable {\it and} superfluid, under realistic experimental conditions.
The resulting metastable phase breaks translation invariance, as the  density profile
averaged over characteristic time scales of interest is inhomogeneous; it
simultaneously  features superfluid order. As mentioned in the introduction,
however, since it lacks density LRO order, the
name ``superglass'' (SG) is more appropriate than ``supersolid", for such a phase.
At the phenomenological level, one may view the
SG  as a special limit of Tisza's two-fluid picture, wherein the normal
fluid component freezes into a glass, instead of forming a quasiparticle
gas.

The first observation of SG was reported in numerical Quantum Monte Carlo
simulations of solid $^4$He samples, quenched from the
normal liquid state at relatively high\footnote{The reader is reminded that the density of
the solid at the $T$= 0 melting line is
$\rho_{m}$ = 0.0287 \Am3} temperature and density $\bar\rho$ = 0.0359 \Am3, to
low temperature ($\sim$ 0.2 K) \cite{B1}. A similar outcome is achieved when samples
are prepared by squeezing the low-temperature superfluid liquid into a
smaller volume. When only local updates of atomic trajectories are employed
in simulations, thus prepared samples of about 800 atoms remain in the
SG phase with inhomogeneous density profile and no density LRO.
Though a direct connection with experiments is hard to make with
regards to the sample preparation protocols, it seems that broad metastability
limits for the SG state of solid $^4$He are guaranteed by simulations.

Further support to the SG concept, comes from the possibility of
``designing'' a quantum system which is guaranteed to have a SG
as an excited state \cite{Biroli}. The idea is based on two key observations:\\
1. {Given a many-body ``Jastrow" wave function for a system of bosons, namely an
expression of the form (the notation is standard)
\beq
\Psi^{(J)}_G = \prod_{i<j}\ {\rm exp}\biggl [-\frac{1}{2}u(r_{ij})\biggr ],
\label{RC}\eeq
$u$ being an arbitrary function of the distance $r_{ij}\equiv |{\bf r}_i-{\bf r}_j|$
between two particles,
one can construct a Hamiltonian $H_J$, with contact two- and three-body interactions,
for which  $\Psi^{(J)}_G$ is the exact ground state wave function. One may  check by direct
substitution
that $\Psi^{(J)}_G$ is a solution of the Schr\"{o}dinger equation for the many-body Hamiltonian
\beq
H_J =\sum_{i=1}^{N} \frac{{\mathbf p}_i^2}{2m_i} + U({\mathbf r}_1, \dots ,{\mathbf r}_N)
\label{HJ}
\eeq
with
\beq
U=-\frac{\hbar^2}{2m}\biggl [ \sum_{i< j} \nabla^2  u(r_{ij})
- \sum_{i<j, i< k}  \nabla u(r_{ij})\cdot
\nabla u(r_{ik})\biggr ]
\eeq
with energy eigenvalue equal to zero (the spectrum of (\ref{HJ}) is non-negative).}\\
2.{ The square of the Jastrow wave function can be
straightforwardly interpreted as the Boltzmann-Gibbs statistical weight for a
classical system with pair-wise potential ${\cal V}=uT$.
This correspondence extends to all eigenstates
which are common for the Hamiltonian (\ref{HJ}) and the Fokker-Planck operator
governing the evolution of the classical probability distribution
within the framework of the stochastic Langevin type dynamics
\begin{equation}
m_i \frac{{d\mathbf r}_i}{dt} =
-\frac{1}{2}\sum_{i < j} \nabla_i u({\mathbf r}_i -{\mathbf r}_j) + \eta_i(t),
\label{Langevin}
\end{equation}
where $\eta_i(t)$ is the thermal Gaussian white noise characterized by the correlation function $\langle \eta_i^{\alpha}(t) \eta_i^{\beta}(t') \rangle =
m_i \delta_{ij} \delta_{\alpha \beta} \delta (t-t')$. We omit here
the derivation which can be found in standard statistical mechanics text, see Ref.~\cite{Biroli} for details.}

The Jastrow wave function
can describe a supersolid, for the following reasons: {\it a}) its square is
isomorphic to a classical distribution, and classical many-particle systems
have a crystalline ordered state for specific interactions and density. \\
{\it b}) it can be shown to correspond to a quantum many-particle state with a finite condensate fraction.

Quantum-to-classical correspondence can be used to compute both static
and dynamic correlation functions at zero temperature, by
using known classical counterpart results.
This observation directly links the well studied jamming phenomenon in
classical systems, to the long-lived metastable
amorphous density profiles in the quantum case.
As far as superfluidity is concerned, it is guaranteed to exist by
construction.
Admittedly, there are serious drawbacks in the theoretical
construction (\ref{HJ}), originating from pathological properties
of the resulting Hamiltonian $H_J$. This Hamiltonian always has the same
ground state energy eigenvalue (zero), regardless of the system density
(which implies an infinite compressibility), a quadratic dispersion relation
for elementary excitations, and a SG {\it only}
appears as an excitation above a supersolid ground state.
Nevertheless, the above argument furnishes a proof of principle of the
existence of SG, and can be made more physical by adding weak standard
pair-wise potentials to Eq.~(\ref{HJ}), with the expectation that the
overall picture may not change.

\subsection{Shevchenko state}

\begin{figure}[tbp]
\centering
\includegraphics[ width=3.0in]{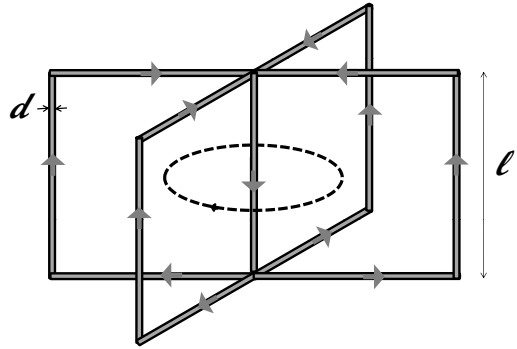}
\caption{The minimal phase defect in the network is obtained by imagining
a vortex ring winding around the pipe and creating $\pm 2 \pi$
phase windings in elementary plaquettes.
}
\label{fig:3}
\end{figure}

At the time of this writing, the most promising theoretical proposal
accounting for transport phenomena observed by
Ray and Hallock in solid $^4$He, seems to be the 3D network formed
by interconnected one-dimensional superfluid channels. Let us start
the discussion with a fictitious system of pipes running along
bonds of a simple cubic lattice, see Fig.~\ref{fig:3}.
Assume that pipes are of diameter
$d$=1 mm, filled with low-temperature ($T \ll T_{\lambda}$)
Helium-II, and the bond length $L$ is such that $ L \gg d^2/a$ where
$a$ is the interatomic distance, say $L=10^6$ km.
The reader might think that the
thermodynamic transition temperature to the SF phase, in
this quasi one-dimensional setup, should be strongly suppressed,
relative to the $\lambda$-point in bulk Helium-II.
However, it is obvious that at, say,  $T < T_{\lambda}/2$,
this network will support frictionless, low-velocity flows, with
persistent currents whose characteristic decay times exceed the age
the Universe! Indeed, at this temperature Helium-II already has
strong local order (definitely on a millimeter scale) in the phase
field, which prevents large vortex excitations from nucleating and
proliferating in the space occupied by helium.

Both expectations are correct, and not in conflict
with each other. Simply, one is dealing here with an ``extreme"
case,  wherein one kinetic time scale rapidly becomes so large, that
in an actual experiment the {\it ergodic hypothesis}, upon which all
equilibrium thermodynamic ensemble calculations are based,
is violated. To make this statement more quantitative, consider the kinetic energy
of a current-carrying state corresponding to a phase winding worth $2 \pi$ around one elementary
plaquette, regarded as isolated from the rest of the system.
A simple calculation based on the integration of the energy density
$\epsilon =(\rho_S/2m)(\nabla \varphi)^2 = (\rho_S/m)(\pi^2 /8L^2)$ gives
\begin{equation}
E_{\rm pl}=\frac{\rho_S a \pi^3}{8m}\: \frac{d^2}{La} \approx T_{\lambda} \: \frac{d^2}{La} \;,
\label{SCH1}
\end{equation}
which is much smaller than $T_{\lambda}$ for $ L \gg d^2/a$
(for parameters mentioned above this energy is in the micro Kelvin range).
In the 3D network, currents are not confined to elementary plaquettes.
The minimal phase defect which can be imagined is a small ``vortex ring''
type configuration depicted in Fig.~\ref{fig:3}.
Though vortexes are not even defined on length scales smaller than
$ \sim L$, and the best microscopic description is provided by phase
gradients along bonds, the topology of phase windings is still easiest to picture using
hypothetical vortex lines. Nevertheless, Eq.~(\ref{SCH1}) sets the scale for the
energy cost of circulating currents in the network. As long as
$E_{\rm pl} \ll T$ the thermodynamic equilibrium state remains normal
because on all scales the closed-contour integrals
$\oint (\nabla \varphi) \cdot d{\bf r}$ are taking non-zero values.

The normal-to-superfluid transition temperature can be estimated from
the condition $E_{\rm pl} \sim T$, at which point plaquette currents become
thermodynamically unfavorable. This condition leads to
\begin{equation}
T_c \sim  T_{\lambda} \: \frac{d^2}{La} \propto \frac{1}{L} \;\;\; (L \gg d^2/a)\;.
\label{SCH2}
\end{equation}
The same estimate for $T_c$ follows from the condition that
1D phase fluctuations on the length-scale $L$ are
reduced to a value of order unity: by introducing an
effective one-dimensional superfluid stiffness $\Lambda^{(1D)} \sim \rho_S d^2/m $
we find that $\langle (\varphi(0)-\varphi(l))^2 \rangle \sim TL/\Lambda_s$.
Once phase fluctuations along the pipe length become small
the three-dimensional order in the network sets in.
The universality class of the transition is not altered by the quasi
one-dimensional microscopic geometry; in particular, the
$\rho_S(T)$ curve starts with an infinite derivative at $T_c$.
By counting the amount of superfluid liquid in the pipes we also find
$\rho_S(T=0) \sim 1/L^2 \sim T_c^2$.

Clearly, the thermodynamic transition discussed above is of purely academic
interest, i.e., one does not expect to observe it experimentally.
At temperatures below $T_{\lambda}$, the notion of thermal equilibrium
quickly becomes irrelevant because any given distribution of persistent
currents is kinetically frozen and cannot change in response to
temperature variations or slow rotation. In other words, in terms of observed behavior,
the state is indistinguishable from the genuine superfluid, despite the fact
that typically one finds large circulating currents on most plaquettes.
This is the essence of the state introduced by S. Shevchenko
\cite{Shevchenko87} to describe the possible superfluid properties of a
dislocation network in the temperature interval
$T_c \ll T \ll T_{\lambda}$, under the hypothesis that edge dislocations
in $^4$He feature superfluid cores.

An important quantitative difference between the fictitious
``pipelines'' example and dislocation network, is that in the latter the pipe diameter
is of the order of the interatomic distance, $d \sim a$.
The condition $L\gg d^2/a$ is modified into $L\gg a$, and thus is
easily satisfied under realistic experimental conditions except,
may be, in a glassy phase. Correspondingly, one may expect that
$T_c \ll T_{\lambda}$, i.e., that there should be a broad temperature interval where the Shevchenko
state may be realized. Another distinction is that narrow superfluid
pipes at low-temperature should be described by the Luttinger liquid theory
with relatively small values of $K$, i.e., their superfluid response is
rather fragile, with stability of circulating currents being no longer protected
by the robust superfluid order on large length scales across the channel.
Long relaxation times, $\tau_{LL}(T)$, for equilibration/redistribution
of plaquette currents leading to the Shevchenko state within
the Luttinger liquid theory are described by the power-law dependence
(not mentioning back-scattering matrix elements leading to momentum
non-conservation)
\begin{equation}
\tau_{LL}(T)\propto \left( \frac{T_{*}}{T} \right) ^{2K-1} \;,
\label{SCH3}
\end{equation}
where $T_*\sim mc^2$ is the characteristic energy scale based on the one-dimensional
sound velocity $c$ below which the Luttinger liquid behavior sets in
\cite{Podlivaev}. For
dislocations in solid helium $T_* \sim T_{\lambda}$.
At temperatures $T\sim T_{\lambda}$, one finds relatively
short relaxation times ensuring normal
state behavior at experimental time scales $\tau_{\rm exper}$. However, at low
temperatures, say for $K \approx 5$ found for screw dislocations, it is
possible to observe a dynamic crossover to the superfluid state when
$\tau_{LL}(T) \gg \tau_{\rm exper}$.
The crossover might be rather sharp and easily accessible
experimentally for large values of $K$.
Note also that typical values of the Luttinger parameter
for ultra-cold atomic systems in the absence of the optical lattice are enormous,
of the order of a hundred or larger ! The question of dynamic crossover
did not arise for the fictitious ``helium pipelines'' system because
the corresponding Luttinger parameter is proportional to the
channel diameter squared $K \sim (d/a)^2$ and the phase-slip
dynamics is frozen right below the $\lambda$-point.

\section{Supersolidity in ultracold atomic systems}
Regardless of how the current controversy over the interpretation of the present $^4$He experiments is eventually resolved, it seems fair to state that solid helium does not afford a direct, simple, and clear observation of the supersolid phenomenon. The question then arises of which other physical system may allow one to make a relatively easy, unambiguous experimental identification of this novel phase.

Among all simple atomic or molecular condensed matter systems, helium offers by far the most favorable combination of large quantum delocalization of its constituent (Bose) particles, owing to the light mass of its atoms, and weakness of the interatomic potential. It is precisely for this reason that liquid helium escapes crystallization at low temperature, under the pressure of its own vapor. The closest condensed matter system that may enjoy similar properties, is molecular hydrogen (H$_2$), also an assembly of Bose particles. Indeed, the mass of a H$_2$ molecule is one half that of a helium atom, which would lead one to expect even higher quantum effects. However, the attractive well of the interaction between two hydrogen molecules is about three times deeper than that of two helium atoms. As a result, liquid hydrogen crystallizes at a relatively high (14 K) temperature, significantly above that at which BEC and SF are expected to occur; although quantum effects, including those of Bose statistics, are detectable in the momentum distribution of the liquid near melting \cite{bon09}, and although superfluid (and even supersolid) behavior has been predicted for small clusters of parahydrogen \cite{mezzacapo1,mezzacapo2,mezzacapo3},  in general, the behavior of solid molecular hydrogen is much closer to that of a classical crystal than to solid helium.  No experimental evidence  for possible superfluid behavior of solid hydrogen has so far been reported \cite{clarkh2}.
\\ \indent
The behavior predicted for solid $^4$He, as it emerges from first-principle quantum simulations, is largely determined by the strong repulsive core of the interatomic potential
at short distance (less than $\sim$ 2 \AA). While the attractive (Van der Waals)  long-range part of the interaction is responsible for the existence  of the condensed phase, it is the repulsive core  (which is a result of Pauli exclusion principle, acting so as to prevent electronic clouds of different atoms from overlapping) that determines most of the thermodynamic properties of condensed helium and other quantum solids and liquids.
For example, a very simple model of Bose hard spheres reproduces surprisingly accurately the phase diagram of condensed helium.
\\
Computer simulation studies of classical crystals, making use of the Lennard-Jones potential, have yielded evidence of the same  vacancy phase separation observed in the quantum system \cite{ma}.  This suggests that the origin of the thermodynamic instability of a gas of point defects lies in the strong interaction among particles, which quantum delocalization cannot overcome.  One is therefore led to consider systems characterized by a different type of pair-wise interaction, possibly with a ``softer" core at short distance. The question is, of course: Where, in nature, does such an interaction arise? The basic features of the helium interatomic potential, chiefly  the strong repulsion at short distance, are common to any molecular or atomic interaction.
\\
One route consists of searching for the supersolid phase in systems in which the ``elementary" constituents are composite particles, e.g., Cooper pairs in superconducting Josephson junction arrays \cite{stroud}, or excitons in electron-hole bilayers \cite{blabla}. In this case, particles dynamically form and disappear due to  pair breaking and recombination effects; the interaction between two such objects is of the effective kind, i.e., induced by the medium in which these particles are embedded, and has an important time-dependent  component. On the other hand, if one wishes to retain the simple picture of ``elementary" particles interacting via a static pair potential, dilute assemblies of spatially confined ultracold atoms appear to offer a viable option.
\\ \indent
Seventeen years after the first successful observation of BEC in a spatially confined assembly of Rubidium atoms, cooled down to temperature in the nanoKelvin range \cite{tutti,pethick}, impressive scientific and technological advances have made  the field of ultracold atoms the natural playground, where fundamental issues in condensed matter and many-body systems can be addressed.
Dilute assemblies of cold atoms constitute an almost ideal many-body system {\it a}) virtually free from the imperfections and ``background noise" that mask subtle physical effects in a real solid, and  {\it b}) upon which a remarkable degree of control can be achieved. In particular, there exist a number of techniques allowing one to vary the interaction between atoms or molecules, rendering it essentially an adjustable parameter. The simplest, and so far most commonly utilized such technique, takes advantage of  the so-called {\it Feshbach resonance} (see, for instance, \textcite{pethick}), whereby the strength of a (short-ranged) interaction between two atoms or molecules can be varied, and even its sign reversed (i.e., turned from repulsive to attractive, or vice versa).
\\ \indent
``Fashioning" artificial interparticle potentials, not arising in any known condensed matter system, allows one to address a key theoretical question, namely which two-body interaction potential(s), if any, can lead to the occurrence of a supersolid phase.
\\
In the next two subsections we shall discuss two physical systems, both realizable in the laboratory using
cold atoms or molecules, which may provide a direct pathway to the stabilization and presumably
straightforward observation of a supersolid phase, owing to the particular interactions among elementary constituents.

\subsection{Rydberg Blockade and soft-core potentials}
The{\it Rydberg blockade} is a physical mechanism that was initially introduced as a way to manipulate quantum information stored in collective states of mesoscopic ensembles \cite{lukin}, has been recently proposed as a way to engineer a novel type of interaction potential  between cold atoms. Specifically, the modified interaction ``flattens off", and  remains essentially constant below some characteristic ``cut-off" distance $a$ \cite{henkel10}. In particular, it is $v(r\to 0) = V$, finite and not much larger than the characteristic energies at play\footnote{For comparison, one may note that the most accurate model potential between two helium atoms \cite{aziz79} is also finite in the $r\to0$ limit, but its value, of the order of $10^6$ K, greatly exceeds any energy scale relevant
to ordinary condensed matter physics.}.
\begin{figure}[tbp]
\centering
\includegraphics[ width=3.0in]{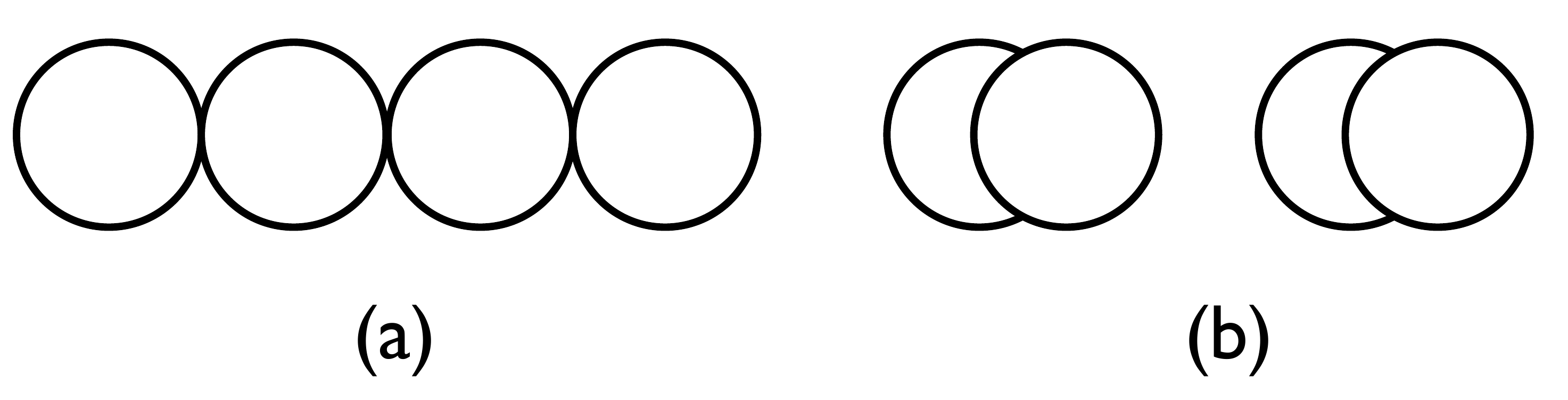}
\caption{One-dimensional crystal of particles interacting via a soft-sphere potential, equal to some energy $V > 0$ if the distance between two particles is less than their effective diameter $a$, and zero if it is greater. As the density equals 1/$a$, configuration (b) becomes energetically advantageous over (a), as its cost is $V$/2 per particle, as opposed to $V$ as in (a).
}
\label{fm1}
\end{figure}

What is the connection between such a potential and the supersolid phase ?
\\
Within the classical complex field description based on the
Pitaevskii-Gross equation, the answer was first provided by E. Gross more than 
half a century ago \cite{Gross}, and more recently quantified  
by Josserand {\it et al.} \cite{josserand} who observed that at sufficiently high
density, the system will break translation invariance and develop solid LRO.
As in any other classical field model, the ground state is
necessarily superfluid, since the lowest energy configuration
corresponds to perfectly ordered phases. Note that the notion of a
"particle" is completely lost within the classical field description,
and thus this study corresponds to the limit of infinite number of particles
(and vacancies) per unit cell. Such a treatment is tantamount to
regarding a supersolid as a superfluid with a density modulation.
Similar results were obtained in Ref.~\cite{henkel10}.

Consider for definiteness a simple model in which particles behave as
soft spheres, namely interact via a ``box" potential, equal to some
energy $V > 0$ if two particles are at a distance $r$ from each other less
than their effective diameter $a$, and zero otherwise.
These above considerations suggest that, if  $V$ is small, and the
number of particles per unit cell is large,
a supersolid phase will be stable if the dimensionless coupling
parameter $g = (Vma^2/\hbar^2)(\bar \rho a^3)$ is greater than unity
(not too large, or the system will behave essentially as a classical crystal).
At the microscopic level, one can intuitively understand why, as a consequence of the constance of the potential at short distance, at sufficiently high density the system will find it energetically favorable to form crystals  with relatively large numbers of particles per unit cell.
\\
As a simple illustration, consider the one-dimensional crystal shown in Fig. \ref{fm1}(a). Assume
again the simple soft-sphere model described above.
Clearly, at a density 1/$a$ each particle can be made to
touch just two others, as shown in Fig. \ref{fm1}(a), at a total energy cost of $V$ per particle. The system can lower its potential energy, by slightly displacing every other particle (i.e., moving pairs of particles closer together) and creating a crystal of lattice constant 2$a$, with two particles per unit cell, as shown in Fig. \ref{fm1}(b).
As the density is increased, the number of particles per units cell also increases.  \\
The outstanding questions, {\it vis-a-vis} the predictions made in Refs. \cite{josserand,henkel10}, are {\it a}) how large a number of particles per unit cell one might require in practice, and {\it b}) what would be the
physical nature of such a supersolid phase.

\begin{figure}[tbp]
\centering
\includegraphics[ width=3.3in]{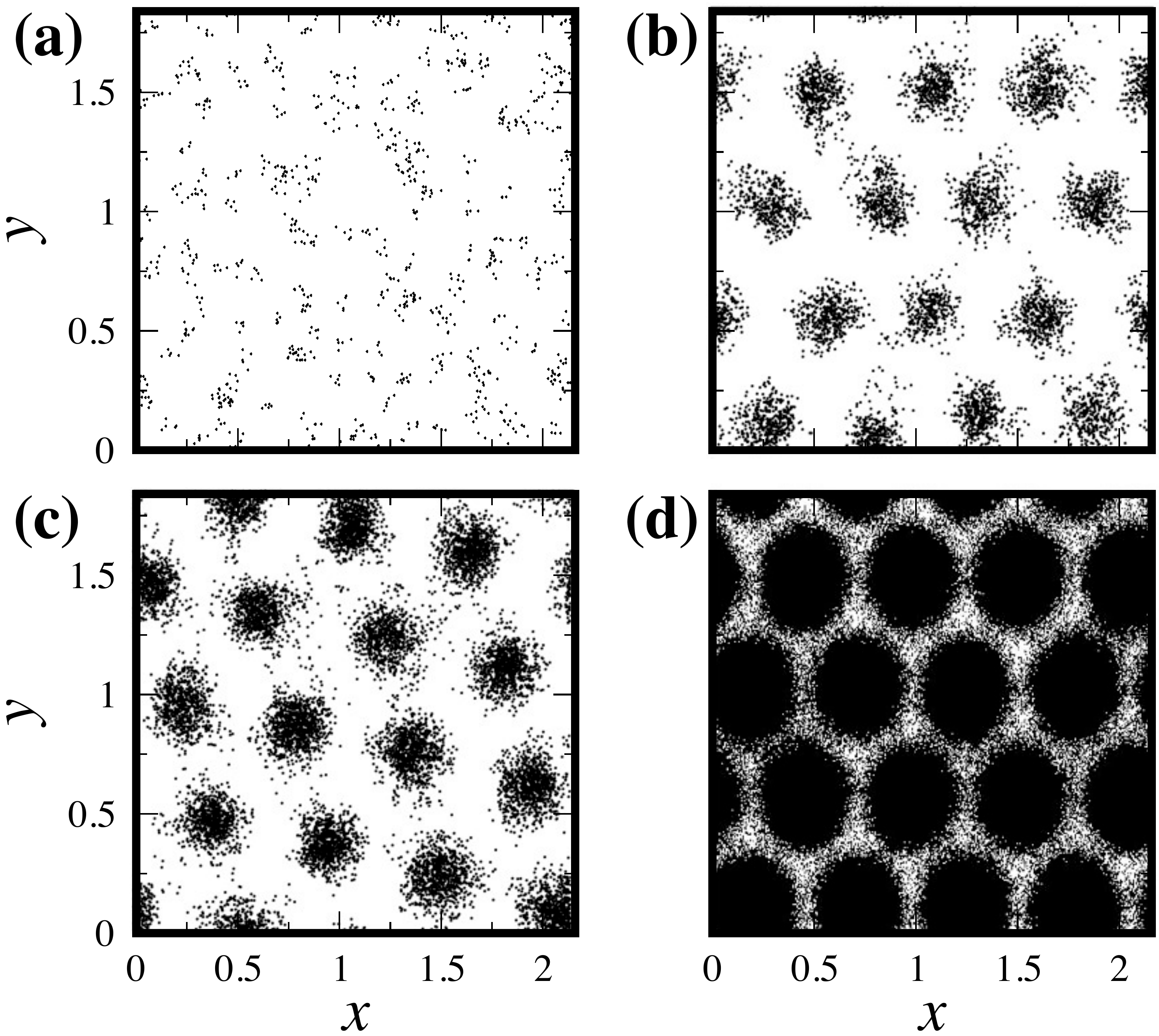}
\caption{Snapshots from computer simulations of a system of bosons  interacting via a power-law potential flattening off at short distances (see Ref. \cite{cinti} for details). Snapshots are taken at four different temperatures, decreasing from (a) to (d);  each snapshot is taken at a temperature an order of magnitude lower than the immediately previous one. Reprinted from Ref. \cite{cinti}.
}
\label{fm2}
\end{figure}

In order to have a fully quantum mechanical quantitative solution of the problem
Cinti {\it et al.} \cite{cinti}, investigated by Quantum Monte Carlo simulations the low temperature phase diagram of a two-dimensional assembly of Bose particles, interacting via a number of different potentials of the general form $v(r) \sim 1/r^n$ for $r > a$, but either equal to a constant or slowly varying for $r <a$, in any case smoothly approaching a finite value for $r \to 0$. In those studies, $n$ was taken equal to 3 and 6. As it turns out,  the main physical results are independent on the behavior of the potential for $r > a$. Indeed, recent work by Saccani {\it et al.} shows that the simple model of ``soft spheres" described above, in two dimensions, features the same basic physics outlined below \cite{saccani}.

A typical result is shown in Fig. \ref{fm2}. Displayed are snapshots of
Feynman's trajectories in imaginary time from a computer simulation. In the particular case shown, the average interparticle distance is approximately $a/2$, and the values of the temperature (which decreases from (a) to (d)) are such that each is an order of magnitude higher (lower) than the next (previous) one (see Ref. \cite{cinti} for details).  As $T$ is decreased, particles bunch into mesoscopic droplets, in turn forming a regular (triangular) crystal. We henceforth refer to this phase as the \textit{droplet-crystal} phase.
The formation of such droplets is a purely classical effect; indeed,
simple (but remarkably accurate) estimates of the average number  $N_d$ of particles per droplet, can be obtained based on purely
classical potential energy minimization, for the various potentials considered.

In the $T\to0$ limit, long exchanges of identical particles can take place, as a result of particles tunneling from one droplet to an adjacent one. In turn, long exchanges of particles can result in a finite superfluid response throughout the whole system, and indeed a {\it bulk} superfluid signal is observed at low temperature, roughly in correspondence of an  inter-particle distance $r_s$ of the order of (slightly less than) the potential cutoff radius. The number $N_d$ of  particles per droplet for which a supersolid phase is observed,  is variable, but can be as low as $\sim$ 4, for specific choices of $r_s$ and $a$. Because superfluidity arises in concomitance with the droplet-crystal structure, the denomination {\em supersolid} seems indisputable in this case. Supersolid behavior in this system originates from tunneling of particles between droplets which are themselves individually superfluid, as simulation results show. This is reminiscent of the phase-locking mechanism in a (self-assembled) array of Josephson junctions.

\subsection{Dipolar systems}
The supersolid phase described in the previous section is a direct consequence of the ``flatness" of the potential at short distance, while the long-range behavior is largely irrelevant. On the other hand, supersolid behavior can also be underlain by long-ranged interactions. We consider here the case of  atoms or molecules possessing a finite electric dipole moment. These particles can be confined to quasi-2D,  by means of an external harmonic potential in the direction perpendicular to the motion (the so-called ``pancake" geometry). Upon aligning all dipoles in the direction perpendicular to the plane, by means of a strong electric field, one can study a system of Bosons interacting  via a purely repulsive potential \cite{minkia} of the form $1/r^3$.

What renders such a system particularly intriguing, is the existence of
an exact theoretical result \cite{Spivak},  excluding the occurrence of
first-order phase transitions involving a density
change $\delta n$ (such as liquid-solid),
in 2D systems with such an interaction.
The reasoning goes as follows: a first-order phase transition is
characterized by the coexistence of two phases of different density,
separated by a macroscopic interface. However, a straightforward calculation
shows that the energy of such an interface contains a negative term,
which diverges logarithmically in the thermodynamic limit.

An immediate consequence of the above result is that,
on approaching the transition  from the low-density (liquid) phase,
the system will lower its free energy by embedding sufficiently
large solid domains (i.e., macroscopic
``bubbles") inside the liquid. At $T=0$ two effects are expected
to occur, namely {\it a}) the transition of the liquid to a superfluid, and {\it b}) the crystallization of solid bubbles into a  lattice superstructure, resulting
in a  global supersolid phase (in fact, a whole set of
different such phases \cite{Spivak}).

An intrinsic subtlety of this scenario is the competition between
the positive contribution to the surface tension, $\sigma_+$,
originating from short-range physics, and the negative scale-dependent
contribution $\sigma_-(R) \propto -(\delta n)^2 \ln (R \rho^{1/2})$,
where $R$ is the droplet size. For small density differences (and
quantum Monte Carlo seem to indicate that this is indeed the case)
the relevant length scale for having negative surface tension
$\sigma (R) = \sigma_+ + \sigma_-$ might be astronomically large
and outside the reach of realistic experimental/numerical setups.

\section{Conclusion}
The investigation of the possible supersolid phase of matter in helium is ongoing, and at the time of this writing no single theoretical framework has emerged as the accepted  interpretation of the puzzling and controversial phenomenology. In this Colloquium, we have presented one possible scenario, which is based on the superfluid properties of extended defects, chiefly dislocations. This scenario is based on first principle microscopic calculations, whose main quality, in our view, is the lack of any {\it a priori} assumption on the physical behavior of the system. It seems consistent with at least an important part of the phenomenology, while attempting to establish a connection between it, and microscopic mechanisms. Whether it will  stand the test of time hinges on its ability to offer quantitatively more accurate microscopic predictions, e.g., on the supersolid transition temperature, {\it if} it is ultimately established that what has been observed in the laboratory, is indeed a signature of superfluidity of the helium crystal. It should be mentioned that, since its introduction, there has been further, independent theoretical work on the Shevchenko state (or, very similar theoretical pictures) by other authors who 
adopted a different, macroscopic approach \cite{toner08,goswami}.

We have also discussed the possibility of realizing the supersolid phase in a different context, namely cold atoms. This is motivated by our belief that such a physical setting may well afford the unambiguous experimental observation of  this intriguing phase, more directly (and with less controversy) than in solid helium.

\section*{Acknowledgments}
This  work was supported in part by the Natural Science and Engineering Research Council of Canada under Research Grant No. 121210893, and by the National Science Foundation under Research Grant PHY-1005543.

\bibliography{references}

\end{document}